\documentclass[10pt]{amsproc}
\input epsf.tex
\newtheorem{theorem}{Theorem}[section]

\theoremstyle{definition}

\newtheorem{example}[theorem]{Example}

\theoremstyle{remark}
\newtheorem{remark}[theorem]{Remark}
\newtheorem{proposition}[theorem]{Proposition}
\def\one{1 \!\! 1}
\numberwithin{equation}{section}

\newcommand{\blankbox}[2]{\parbox{\columnwidth}{\centering}}
\input{diagrams.tex}

\begin{document}

\title{On quantum symmetries of ADE graphs}

\author{Robert Coquereaux}
\address{Centre de Physique Th\'eorique, Luminy, 13288 Marseille, France}
\email{coque@cpt.univ-mrs.fr}
\author{Roberto Trinchero}
\address{Centro At\'omico Bariloche e Instituto Balseiro,
         8400 Bariloche, Argentina.}
\email{trincher@cab.cnea.gov.ar}
\address{R. C. is supported by CNRS, and R. T. by CONICET}
%\date{}

\begin{abstract}
The double triangle algebra(DTA) associated to an ADE graph is considered. A description of its bialgebra structure based on a reconstruction approach is given. This approach takes as initial data the representation theory of the DTA as given by Ocneanu's cell calculus. It is also proved that the resulting DTA has the structure of a weak *-Hopf algebra. As an illustrative example, the case of the graph $A_3$ is described in detail. 
\end{abstract}

\maketitle

\newarrow{Line}{-}{-}{-}{-}{-}
\newarrow{Iline}{=}{=}{=}{=}{=}
\newarrow{Dline}{}{dash}{}{dash}{}
\section{Introduction}

This paper deals with the correspondence between rational conformal field theories(RCFT) of $SU(2)$-type and $ADE$ graphs \cite{cappelli, CIZ, Pasquier}.
More precisely we focus on the construction of the so-called double triangle algebra(DTA)\cite{Ocneanu:paths}  associated to an ADE graph. The DTA is a bialgebra and the  ``algebra of quantum symmetries''  describes the tensor product of representations associated with one of its product structures (the same name sometimes denotes the bialgebra itself).
Aside from its interest as a mathematical structure, the motivation in considering the DTA stems from the fact that its knowledge makes it possible to construct the modular invariant partition function, as well as other objects, in particular the so-called twisted partition functions \cite{Coque:Qtetra, CoqueGil:ADE, robmar, pz, Gil:thesis}  associated with the corresponding RCFT -- actually \cite{pz} one has to take into account existence of boundaries and defects.

The present work contributes to the understanding of the DTA in two respects. First, it  provides a precise description of the DTA. To the knowledge of the authors such a description is not available in the literature. Furthermore our approach is constructive in the sense that, given an ADE graph,
one can construct the corresponding DTA{\footnote{The main computational effort is to compute the connections associated with the corresponding Ocneanu cell systems.}}.
This is done starting from the calculus of Ocneanu's cells and connections. More precisely, the approach we employ amounts essentially to take the above mentioned calculus as describing the representation theory of an algebraic structure to be found: the DTA.
In this paper\footnote{A very simple example is analysed in \cite{Coque:CocoyocA2}  but notations are not the same (and the point of view is quite different).}, the DTA has a product  called $\cdot$ which  is determined 
from the cell calculus. It  has also a coproduct $\Delta$. This coproduct determines a product on the dual $\widehat{DTA}$ that corresponds to the composition of endomorphisms.
In other references this later product  is called "composition product"
whereas the product $\cdot$ that we study here is called "convolution product". We decided to focus  the present paper on the convolution product since this operation is the non-trivial one\footnote{These properties could  be discussed in terms of nets of subfactors \cite{Evans, Evans:Bariloche}, a notion that  we do not use here.}: the other operation is simply the composition of endomorphisms stemming from the definition of the underlying vector space structure of the DTA (this vector space is defined as  $End^{gr}(\mathcal E)$, the graded vector space $\mathcal E$ will be defined later). 
 
The second subject we address is the assertion that the DTA has the structure of a weak *-Hopf algebra (WHA) \cite{bs}. This assertion is not new, it is given in ref.\cite{pz} that gives arguments based on considering solutions of the so-called \cite{bs, Ocneanu:StFrancois}  "big pentagon equation"(BPE).
In contrast to  ref.\cite{pz} we do not assume any a priori knowledge of such a solution. 
In our work, each structural map of the corresponding WHA is  constructed 
in terms of the available data, i.e. connections on cell systems, and all
the WHA properties of these maps are proved using properties of the cell calculus (which was introduced in \cite{Ocneanu:paragroups} and described for instance in \cite{EvansKawa:book} and \cite{Roche}).
 In order to make contact with ref.\cite{pz}, it would be desirable to 
establish a precise connection between the present results and solutions of the BPE .

The paper is organized as follows. Section II  gives  generalities about graphs and specify the requirements we want the DTA to fulfill. These requirements deal mainly with the representation theory of a "searched for" C$^*$-algebra, that we call DTA. The remaining sections uncover the
structure maps of the DTA out of the data given  in section II. Section III deals with the algebra structure taking advantage of the fact that the DTA is a finite dimensional C$^*$-algebra. Section IV gives in terms of connections, what we may call "the weak bialgebra structure maps of the DTA", that is, product, coproduct and counit in terms of connections. Section V considers the antipode. The main sections are supplemented by four appendices.
All the general results are exemplified in detail for the case of the graph $A_3$.

\section{The double triangle algebra}
\subsection{Preliminaries}
\label{s2.1}
Let us consider a graph $G$ with $n_v$ vertices.
%A bipartite graph is a graph where it
%is possible to separate its vertices in two sets such that there
%are no edges  between vertices belonging to the same set.
One can characterize a graph by its adjacency matrix $M$. Its size is  $n_v \times n_v$ and its $(v_1 , v_2)$ matrix element is
an integer $n$ if vertex $v_1$ is connected to vertex $v_2$ by
$n$ edges. The normalized{\footnote{Set a smallest component to be equal to $1$.}} eigenvector with maximum eigenvalue $\beta$ of the adjacency matrix $M$ is called the Perron-Frobenius eigenvector
and its components will be denoted by $\mu_{v_i}, \;\;i=1, \cdots , n_v$.

We can define over $G$ a vector space $\mathcal P$ whose elements
are paths. An elementary path of length $n$ is a
ordered $n$-uple of contiguous vertices in $G$. Two vertices are
contiguous if there exists an edge connecting them. A path is a
linear combination over $\mathbb C$ of elementary paths. Therefore these elementary
paths provide a preferred basis of $\mathcal P$. 
 
This vector space is graded by the length of paths.
There is a subspace $\mathcal E$ of $\mathcal P$ given by  "essential paths", that is paths
that are annihilated by all  the Ocneanu's operators
$c_k,\;\;k \in \mathbb N$. The operator $c_k$ acting on a path of length $n \leq k$ gives zero, otherwise ($n>k$) it is given  by,
\begin{equation}
\label{1.1}
c_k (v_0, v_1 ,v_2 ,\cdots, v_{k-1}, v_{k}, v_{k+1}, \cdots ,v_n ) =
\sqrt{\frac{\mu_{v_k}}{\mu_{v_{k+1}}} }\delta_{v_{k-1}, v_{k+1}}
(v_0, v_1, \cdots,v_{k-2} , v_{k-1},v_{k+2}, \cdots , v_n)
\end{equation}
where $(v_0, v_1 ,v_2 ,\cdots, v_{k-1}, v_{k}, v_{k+1}, \cdots ,v_n )$
denotes an elementary path of length $n$ passing through the
vertices $v_0$ to $v_n$ of $G$. That is the operator $c_k$
eliminates any one-step backtracking sub-path that starts at step
$k$ of the path to which the operator  is applied and multiplies the result
by a number given in terms of the components of the Perron-Frobenius eigenvector. 
%We consider a scalar product in $\mathcal P$ such that the basis
%of elementary paths is orthonormal.
%This scalar product
%is given by,
%\begin{equation}
%\label{1.2}
%<\xi ,\xi' > = \delta_{\xi \xi'} \qquad .
%\end{equation}

There is a natural product in $\mathcal P$ defined by concatenation of paths. The concatenation product of two paths is zero if the ending vertex of the first path is not equal to the starting vertex of the second  path. If the above holds then the product path is simply the extension of the first path by the second. In symbols
take $\xi_i = (v_0^i , \cdots , v_n^i)$ and 
$\xi_j = (v_0^j , \cdots , v_m^j)$ then the concatenation product $\xi_i \star \xi_j$ of $\xi_i$ and $\xi_j $ is given by, 
\begin{equation}
\label{2}
\xi_i \star \xi_j = \delta_{v_n^i v_0^j} 
(v_0^i , \cdots , v_n^i,v_1^j, \cdots , v_m^j) \qquad .
\end{equation}

We shall restrict to graphs where the dimension of $\mathcal E$ is
finite. This restriction is very strong and essentially\footnote{One could be more accurate, but such a discussion is  not needed for our present purpose. Notice that an extension of the  results described here to other types of graphs, for example to graphs belonging to higher Coxeter-Dynkin systems related to $SU(N)$-type RCFTs \cite{DiF-Zub-Trieste, DiFrancescoZuber, Ocneanu:Bariloche, z, Zuber:generaldynkin}  requires an appropriate modification of the definition of essential paths.}  reduces the family of  admissible graphs to those belonging to the ADE series \cite{Ocneanu:paths}.

The basis of elementary paths restricted to the maximum length of essential path will be denoted by $\{\xi_i \}$.
We are interested in the length preserving endomorphisms of $\mathcal E$ that we
denote $End^{gr}(\mathcal E)$. We denote the dual vector space of $\mathcal E$ by $\hat{\mathcal E}$ and by $\{\xi^i \}$ the dual basis
to the $\{\xi_i \}$ . Hence a basis of $End^{gr}(\mathcal E)$ is given by the objects $\{\xi_i \otimes \xi^j \} $.
 
\begin{example}[The case of $A_3$.]
\label{ex1}
The graph $G=A_3$ and its corresponding adjacency matrix $M$ are,
\begin{equation}
\label{e1}
{
\begin{diagram}[size=0.8em,abut]
0 & \, & 1 & \, & 2 \\
\bullet&\rLine & \bullet &\rLine & \bullet \\
\end{diagram}
}
\qquad , \qquad
M=
\left(
\begin{array}{lll}
 0&1 &0\\
 1&0 &1\\
 0&1 &0
 \end{array}
\right)
\end{equation}
where rows and columns are ordered as $0,1,2$ (the values for vertex $v$ can be $0,1$ or $2$).
The maximum eigenvalue is $\beta=\sqrt{2}$ and the Perron-Frobenius eigenvector is $(1,\sqrt{2},1)$.
There are ten essential paths in this graph. Denoting them by the corresponding succession of vertices, they are the following,
\begin{equation}
\label{e2}
\begin{array}{llll}
(0)=0 & (1)=1 & (2)=2 & \\
(01)=r_0&(12)=r_1 &(10)=l_1  &  (20)=l_2\\
(012)=d& (210)=g &  \frac{(121)-(101)}{\sqrt{2}}=\gamma &
\end{array}
\end{equation} 
where we have also included a shorthand notation. From (\ref{e2}) we see that the dimension of $End^{gr}(\mathcal E)$ is $3^2 +4^2 + 3^2=34$.
\end{example}

\subsection{The double triangle algebra(DTA)}
As mentioned in the introduction the strategy adopted in this paper is to reconstruct an algebraic structure out of certain requirements that we want to be fulfilled. They will involve information about the representation theory but not directly about the product structure.
Indeed one of the results of this analysis will be the  product $\cdot$ for this algebra.   
Requirements:

\begin{enumerate}
\item Vector space structure: It is given by $End^{gr}(\mathcal E)$.

We emphasize that this requirement   is only relative to choice of  the underlying vector space structure and not the product itself since the  product law that we study on the vector space $End^{gr}(\mathcal E)$
is ({\em not }) the composition of endomorphisms but another product that we call $\cdot$. Rather than giving this product explicitly we shall obtain it by describing all its representations. 

\item It is a $C*$-algebra{\footnote{Being a $C^*$-algebra it is 
semisimple, and is therefore graded for this product $\cdot $ (simple blocks) but this grading differs from the one given by the length of paths in $End^{gr}(\mathcal E)$.}}.
\item Fundamental {\footnote{Fundamental in the sense that they generate any other irrep by taking adequate linear combinations of tensor products of them. See item (4) for the definition of tensor product representations.}}irreps of the product $\cdot$: These are matrix $*$-representations {\footnote{By this we mean that the star operation in the DTA corresponds to hermitian conjugation of matrices in these representations}}$f$ 
whose matrix element $\alpha \beta $ is
given by
{\footnote{\label{5}In the drawing (\ref{1.4}) the labels for horizontal paths $\xi$ and $\xi'$
denote a generic element of $\mathcal E$ and $\hat{\mathcal E}$ respectively.}},
\begin{equation} 
\label{1.4}
\Phi^f_{\alpha \beta}( \xi \otimes \xi')=
{
\begin{diagram}[size=0.8em,abut]
\,&\xi  &\, \\
\bullet&\rTo~{n} & \bullet  \\
\dTo^{\alpha}&  f      &\dTo_{\beta} \\
\bullet&\rTo & \bullet  \\
\, &\xi'  &\, \\
\end{diagram}
}
\end{equation}
where the matrix indices $\alpha , \beta$ label   length-one paths
on $G$. If it does not happen that,
\begin{equation}
\label{1.4'}
s(\alpha ) = s(\xi) \;\;, r(\alpha)=s(\xi') \;\;, s(\beta)= r(\xi) \;\;, r(\beta ) = r(\xi')
\end{equation}
then (\ref{1.4}) vanishes{\footnote{In (\ref{1.4'}) we have denoted by $s(\alpha)$ the starting vertex of path $\alpha$ and by $r(\alpha)$ its final vertex.}}.
The symbol on the r.h.s. of (\ref{1.4}),
is a complex number and is by definition the connection associated to that cell in a
fundamental representation.
These values should satisfy the following conditions,
\begin{enumerate}
\item[(0)] Zero length paths: 

If $\# \xi = \#\xi' = 0$ then,
\begin{equation}
\label{1.4''}
{
\begin{diagram}[size=0.8em,abut]
\,&\xi  &\, \\
\bullet&\rTo~{n} & \bullet  \\
\dTo^{\alpha}&  f      &\dTo_{\alpha} \\
\bullet&\rTo & \bullet  \\
\, &\xi'  &\, \\
\end{diagram}
}
= \delta_{s(\alpha) \xi} \delta_{r(\alpha) \xi'}
\end{equation}

\item[(i)] Unitarity{\footnote{Using (\ref{1.5}) and (\ref{1.6}) the following relation is obtained,
\begin{equation}
\label{1.4'''}
\sum_{\beta , \,\xi} 
{\overline
{
\begin{diagram}[size=0.8em,abut]
\,&\xi  &\, \\
\bullet&\rTo~{n} & \bullet  \\
\dTo^{\alpha}&  f      &\dTo_{\beta} \\
\bullet&\rTo & \bullet  \\
\, &\xi'  &\, \\
\end{diagram}
}}\;
{
\begin{diagram}[size=0.8em,abut]
\,&\xi  &\, \\
\bullet&\rTo~{n} & \bullet  \\
\dTo^{\alpha'}&  f      &\dTo_{\beta} \\
\bullet&\rTo & \bullet  \\
\, &\lambda  &\, \\
\end{diagram}
}
=\delta_{\alpha \alpha'} \delta_{\xi' \lambda}
\end{equation}
conditions (\ref{1.5}) and (\ref{1.4'''}) are called "bi"unitarity in ref.\cite{Ocneanu:paths}.
}}:
\begin{equation}
\label{1.5}
\sum_{\alpha , \,\xi'} 
{\overline
{
\begin{diagram}[size=0.8em,abut]
\,&\xi  &\, \\
\bullet&\rTo~{n} & \bullet  \\
\dTo^{\alpha}&  f      &\dTo_{\beta} \\
\bullet&\rTo & \bullet  \\
\, &\xi'  &\, \\
\end{diagram}
}}\;
{
\begin{diagram}[size=0.8em,abut]
\,&\lambda  &\, \\
\bullet&\rTo~{n} & \bullet  \\
\dTo^{\alpha}&  f      &\dTo_{\eta} \\
\bullet&\rTo & \bullet  \\
\, &\xi'  &\, \\
\end{diagram}
}
=\delta_{\beta \eta} \delta_{\xi \lambda}
\end{equation}
\item[(ii)]Reflection:
\begin{equation}
\label{1.6}
{
\begin{diagram}[size=0.8em,abut]
\,&\xi  &\, \\
\bullet&\rTo~{n} & \bullet  \\
\dTo^{\alpha}&  f      &\dTo_{\beta} \\
\bullet&\rTo & \bullet  \\
\, &\xi'  &\, \\
\end{diagram}
}
= 
\sqrt{\frac{\mu_f^{\xi} \mu_i^{\xi'}}{\mu_i^{\xi} \mu_f^{\xi'}}
}\;\;
{\overline
{
\begin{diagram}[size=0.8em,abut]
\,&{\tilde \xi}  &\, \\
\bullet&\rTo~{n} & \bullet  \\
\dTo^{\beta}&  f      &\dTo_{\alpha} \\
\bullet&\rTo & \bullet  \\
\, &{\tilde \xi'}  &\, \\
\end{diagram}
}}\;
=
\sqrt{\frac{\mu_f^{\xi} \mu_i^{\xi'}}{\mu_i^{\xi} \mu_f^{\xi'}}
}\;\;
{\overline
{
\begin{diagram}[size=0.8em,abut]
\,&\xi'  &\, \\
\bullet&\rTo~{n} & \bullet  \\
\dTo^{{\tilde \alpha}}&  f      &\dTo_{{\tilde \beta}} \\
\bullet&\rTo & \bullet  \\
\, &\xi  &\, \\
\end{diagram}
}}\;
\end{equation}
where $\tilde \alpha$ denotes the path that is obtained from $\alpha$
reversing the arrows.
\item[(iii)]Concatenation properties:

\begin{equation}
\label{1.6'}
{
\begin{diagram}[size=0.8em,abut]
\,&\xi_1  \star \xi_2   &\, \\
\bullet&\rTo~{n} & \bullet  \\
\dTo^{\alpha}&  f      &\dTo_{\beta} \\
\bullet&\rTo & \bullet  \\
\, &\xi'_1 \star  \xi'_2 &\, \\
\end{diagram}
}
=
{
\begin{diagram}[size=0.8em,abut]
\,&\xi_1  &\, \\
\bullet&\rTo~{n} & \bullet  \\
\dTo^{\alpha}&  f      &\dTo_{\gamma} \\
\bullet&\rTo & \bullet  \\
\, &\xi'_1  &\, \\
\end{diagram}
}
{
\begin{diagram}[size=0.8em,abut]
\,&\xi_2  &\, \\
\bullet&\rTo~{n} & \bullet  \\
\dTo^{\gamma}&  f      &\dTo_{\beta} \\
\bullet&\rTo & \bullet  \\
\, &\xi'_2  &\, \\
\end{diagram}
}
\end{equation}
where the length one path $\gamma$ is determined by the ending(starting) vertices of $\xi_1$($\xi_2$) and $\xi'_1$($\xi'_2$). 
\begin{remark}
It is very important to realize at this stage that, although connections have been 
defined in eq.(\ref{1.4}) for elements in  $End^{gr}(\mathcal E)$, property (iii)
allows to define them for elements in  $End^{gr}(\mathcal P)$. This is so since by concatenation as in eq.(\ref{1.6'}) it is possible to build any path out of length-zero and length-one paths that are necessarily essential.
\end{remark}
\end{enumerate}
\item The tensor product representation is given by,
\begin{equation}
\label{1.7}
\Phi^{f \otimes f'}_{\alpha \star \beta ,\alpha' \star \beta' } (\xi \otimes \xi' )
=\sum_{\xi_i}
\Phi^{f}_{\alpha  ,\alpha'  } (\xi \otimes \xi^i )
\;
\Phi^{f'}_{\beta  ,\beta'  } ( \xi_i  \otimes \xi')
\qquad . 
\end{equation} 
where $f, f'$ can be any of the fundamentals and where the dually paired basis vectors
$\xi_i$ and $\xi^i$ have been defined at the end of section \ref{s2.1}.
\end{enumerate}
\begin{example}[The case of $A_3$.]
In this case there is only one fundamental representation that according to what is shown in appendix A can be chosen as,
\begin{equation}
\label{1.8}
{
\begin{diagram}[size=0.8em,abut]
0&r_0   &1 \\
\bullet&\rTo & \bullet  \\
\dTo^{r_0}&  f      &\dTo_{l_1} \\
\bullet&\rTo & \bullet  \\
1 &l_1  & 0\\
\end{diagram}
}=1
\;\;,
{
\begin{diagram}[size=0.8em,abut]
0&r_0  &1 \\
\bullet&\rTo & \bullet  \\
\dTo^{r_0}&  f      &\dTo_{r_1} \\
\bullet&\rTo & \bullet  \\
1 &r_1  & 2\\
\end{diagram}
}=1
\;\;,
{
\begin{diagram}[size=0.8em,abut]
2& l_2  &1 \\
\bullet&\rTo & \bullet  \\
\dTo^{l_2}&  f      &\dTo_{l_1} \\
\bullet&\rTo & \bullet  \\
1 &l_1  & 0\\
\end{diagram}
}=1
\;\;,
{
\begin{diagram}[size=0.8em,abut]
2& l_2  &1 \\
\bullet&\rTo & \bullet  \\
\dTo^{l_2}&  f      &\dTo_{r_1} \\
\bullet&\rTo & \bullet  \\
1 &r_1  & 2\\
\end{diagram}
}=-1
\nonumber
\end{equation}
\begin{equation}
\label{1.9}
{
\begin{diagram}[size=0.8em,abut]
1& l_1  &0 \\
\bullet&\rTo & \bullet  \\
\dTo^{l_1}&  f      &\dTo_{r_0} \\
\bullet&\rTo & \bullet  \\
0 &r_0  & 1\\
\end{diagram}
}=1/\sqrt{2}
\;\;,
{
\begin{diagram}[size=0.8em,abut]
1&r_1  &2 \\
\bullet&\rTo & \bullet  \\
\dTo^{l_1}&  f      &\dTo_{l_2} \\
\bullet&\rTo & \bullet  \\
0 &r_0  & 1\\
\end{diagram}
}=1/\sqrt{2}
\;\;,
{
\begin{diagram}[size=0.8em,abut]
1& l_1  &0 \\
\bullet&\rTo & \bullet  \\
\dTo^{r_1}&  f      &\dTo_{r_0} \\
\bullet&\rTo & \bullet  \\
2 &l_2  & 1\\
\end{diagram}
}=1/\sqrt{2}
\;\;,
{
\begin{diagram}[size=0.8em,abut]
1& r_1  &2 \\
\bullet&\rTo & \bullet  \\
\dTo^{r_1}&  f      &\dTo_{l_2} \\
\bullet&\rTo & \bullet  \\
2 &l_2  & 1\\
\end{diagram}
}=-1/\sqrt{2}
\end{equation}
where for the sake of completeness we have included the corresponding vertex labels in each cell.
Using (iii) it is possible to compute the value of cells for longer horizontal paths, for example,
\begin{eqnarray}
\label{1.10}
{
\begin{diagram}[size=0.8em,abut]
0& r_0 \star r_1  &2 \\
\bullet&\rTo & \bullet  \\
\dTo^{r_0}&  f      &\dTo_{l_2} \\
\bullet&\rTo & \bullet  \\
1 & \gamma  & 1\\
\end{diagram}
}
&=& 1/\sqrt{2}
\left(
{
\begin{diagram}[size=0.8em,abut]
0& r_0 \star r_1  &2 \\
\bullet&\rTo & \bullet  \\
\dTo^{r_0}&  f      &\dTo_{l_2} \\
\bullet&\rTo & \bullet  \\
1 & r_1 \star l_2  & 1\\
\end{diagram}
}
-
{
\begin{diagram}[size=0.8em,abut]
0& r_0 \star r_1  &2 \\
\bullet&\rTo & \bullet  \\
\dTo^{r_0}&  f      &\dTo_{l_2} \\
\bullet&\rTo & \bullet  \\
1 & l_1 \star r_0  & 1\\
\end{diagram}
}
\right)\nonumber\\
&=& 1/\sqrt{2} 
\left(
{
\begin{diagram}[size=0.8em,abut]
0&r_0  &1 \\
\bullet&\rTo & \bullet  \\
\dTo^{r_0}&  f      &\dTo_{r_1} \\
\bullet&\rTo & \bullet  \\
1 &r_1  & 2\\
\end{diagram}
}
{
\begin{diagram}[size=0.8em,abut]
1& r_1  &2 \\
\bullet&\rTo & \bullet  \\
\dTo^{r_1}&  f      &\dTo_{r_0} \\
\bullet&\rTo & \bullet  \\
2 &l_2  & 1\\
\end{diagram}
}
-
{
\begin{diagram}[size=0.8em,abut]
0&r_0   &1 \\
\bullet&\rTo & \bullet  \\
\dTo^{r_0}&  f      &\dTo_{l_1} \\
\bullet&\rTo & \bullet  \\
1 &l_1  & 0\\
\end{diagram}
}\;\;
{
\begin{diagram}[size=0.8em,abut]
1&r_1  &2 \\
\bullet&\rTo & \bullet  \\
\dTo^{l_1}&  f      &\dTo_{l_2} \\
\bullet&\rTo & \bullet  \\
0 &r_0  & 1\\
\end{diagram}
}
\right)=1
\end{eqnarray}
which, as we see, corresponds to a horizontal "concatenation" of basic 
cells.
Using 4. we can compute the value of cells in tensor product representation, for example,
\begin{eqnarray}
\label{1.11}
\Phi^{f \otimes f}_{r_0 \star l_1 ,r_1 \star l_2 } (r_0 \otimes r_0 )
&=&
{
\begin{diagram}[size=0.8em,abut]
0&r_0  &1 \\
\bullet&\rTo & \bullet  \\
\dTo^{r_0 \star l_1}&  f \otimes f      &\dTo_{r_1 \star l_2} \\
\bullet&\rTo & \bullet  \\
0 &r_0  & 1\\
\end{diagram}
}
=
{
\begin{diagram}[size=0.8em,abut]
0&r_0   &1 \\
\bullet&\rTo & \bullet  \\
\dTo^{r_0}&  f      &\dTo_{r_1} \\
\bullet&\rTo & \bullet  \\
1 &r_1  & 2\\
\end{diagram}
}
{
\begin{diagram}[size=0.8em,abut]
1&r_1  &2 \\
\bullet&\rTo & \bullet  \\
\dTo^{l_1}&  f      &\dTo_{l_2} \\
\bullet&\rTo & \bullet  \\
0 &r_0  & 1\\
\end{diagram}
}
=1/ \sqrt{2}\nonumber\\
&=&
\Phi^{x}_{r_0  , r_1  } (r_0 \otimes r_1 )
\;
\Phi^{y}_{l_1  ,l_2  } ( r_1  \otimes r_0)
\end{eqnarray}
which, as we see, corresponds to a vertical "concatenation" of 
basic cells.
\end{example}
%%%%%%%%%%%%%%%%%%%%%%%%%%%%%%%%%%%%%%%%%%%%%%%%%%%%%%%%%%%%%%%%%%%%%%%%%%%%%%%%%%%%%%
\section{Algebra structure of the DTA}

\begin{proposition}
\label{p1}
There exists a basis of the DTA denoted by $E^x_{\eta \eta'}$,
where the index $x$ labels the irreducible representations of the
DTA and $\eta, \eta'$ are indices in the irrep $x$. In this basis the
product is given by,
\begin{equation}
\label{2.1}
  E^{x_1}_{\eta_1 \eta'_1} \cdot E^{x_2}_{\eta_2 \eta'_2} =
  \delta_{\eta'_1 \eta_2}\delta_{x_1 x_2} E^{x_1}_{\eta_1 \eta'_2}
\end{equation}
The identity for this product is given by,
\begin{equation}
\label{2.2}
\one = \sum_{x,\eta} E^x_{\eta \eta}
\end{equation}
The matrix *-representations (\ref{1.4}) are homomorphisms $\Phi^x$ in the following way,
\begin{equation}
\label{2.3}
\Phi_{\alpha\gamma}^x 
(E^{x_1}_{\eta_1 \eta'_1}\cdot E^{x_2}_{\eta_2 \eta'_2})=
\sum_{\beta} \Phi_{\alpha\beta}^x 
(E^{x_1}_{\eta_1 \eta'_1})
\Phi_{\beta\gamma}^x 
(E^{x_2}_{\eta_2 \eta'_2})
\end{equation}
 The star structure takes
 the following form,
\begin{equation}\label{2.4}
(E^x_{\eta \eta'})^* = E^x_{\eta' \eta}
\end{equation}
The scalar product is given by,
\begin{equation}
\label{2.4'}
<E^{x_1}_{\eta_1 \eta'_1} |E^{x_2}_{\eta_2 \eta'_2}> = 
\delta_{x_1 x_2} \delta_{\eta_1 \eta_2} \delta_{\eta'_1 \eta'_2}
\end{equation} 
\end{proposition}
\begin{proof}
Since we are restricting to graphs where the dimension of $\mathcal E$ is finite and using 2. we conclude that the DTA we are considering are finite dimensional $C*$-algebras. Hence they are isomorphic to a direct sum of matrix algebras corresponding to their irreps. Since we are dealing with matrix algebras we can take a basis consisting of matrix units,
\begin{equation}
\label{2.5}
(E^x_{\eta \eta'})_{\alpha \beta} = \delta_{\eta \alpha} \delta_{\eta' \beta}
\end{equation}
in this basis of  matrix units, matrix multiplication corresponds to (\ref{2.1}). That $\one$ in eq.(\ref{2.2}) is the identity for this product is very simple to verify.
As we mentioned above, the terms of this direct sum decomposition\footnote{The discussion concerning the direct sum decomposition of tensor product representations can be rephrased in graphical terms by using the notion of cell systems associated with  representations\cite{cst}.} correspond to the irreducible representations of the DTA
\begin{equation}
\label{2.6}
\Phi^y_{\alpha \beta} (E^x_{\eta \eta'})= \delta_{xy} \delta_{\alpha \eta} \delta_{\beta \eta'}
\end{equation}
Replacing (\ref{2.6}) in (\ref{2.3}) you verify that the later holds.
Regarding the star structure using 3. we have,
\begin{equation}
\label{2.7}
\Phi^y_{\alpha \beta} ((E^x_{\eta \eta'})^*)= 
{\overline{\Phi^y_{\beta \alpha} (E^x_{\eta \eta'})}}=
\delta_{xy} \delta_{\beta \eta} \delta_{\alpha \eta'}
=\Phi^y_{\alpha \beta} (E^x_{\eta' \eta})
\qquad \forall x,y,\eta, \eta',\alpha, \beta
\end{equation}
which leads to (\ref{2.4}).
Regarding the scalar product in the basis $\{E^x_{\eta \eta'}\}$ 
we note that since there is a unique correspondence between C$*$-algebras and operator algebras and since the scalar product
(\ref{2.4'}) leads to the operator norm, it must be that one. 
\end{proof}
%We recall the following theorem appearing in 2.7.3 of ref.\cite{d}.
%{\it{The irreducible representations of a C*-algebra are such that %the norm in the C*-algebra corresponds to the operator norm in the %irreducible representations.}}%*****Eventually to a footnote.****

%%%%%%%%%%%%%%%%%%%%%%%%%%%%%%%%%%%%%%%%%%%%%%%%%%%%%%%%%%
%*******************************************************
\subsection{Relation between basis}

Let us adopt the notation that matrix irreps are represented by similar symbols as in the case of the fundamentals, i.e.,
\begin{equation}
\label{2.8}
\Phi^x_{\alpha \beta}( \xi \otimes \xi')=
{
\begin{diagram}[size=0.8em,abut]
\,&\xi  &\, \\
\bullet&\rTo~{n} & \bullet  \\
\dTo^{\alpha}&  x      &\dTo_{\beta} \\
\bullet&\rTo & \bullet  \\
\, &\xi'  &\, \\
\end{diagram}
}
\end{equation}
where $x$ runs over all the irreps of the DTA(not only the fundamentals) as in proposition (\ref{p1}). It is worth noting that condition (\ref{1.4'}) should be fulfilled also in the case where $x$ is not one of the fundamentals. This is so because the fundamentals tensorially generate any other irrep, recalling the definition (\ref{1.7}) of tensor product representation we see that condition (\ref{1.4'}) is also fulfilled in any tensor product of fundamentals.

We have introduced two basis for the DTA, one in terms of endomorphisms of essential paths and another in terms of matrix units for the product in the DTA, there is a relation between them as stated by the following proposition,
\begin{proposition}
\label{p2.3}
The two basis $\{\xi \otimes \xi'\}$ and $\{E^x_{\alpha\beta}\}$ are related by,
\begin{equation}
\label{2.13}
\xi \otimes \xi' =\sum_{x, \eta, \eta'}
{
\begin{diagram}[size=0.8em,abut]
\,&\xi  &\, \\
\bullet&\rTo~{n} & \bullet  \\
\dTo^{\eta}&  x      &\dTo_{\eta'} \\
\bullet&\rTo & \bullet  \\
\, &\xi'  &\, \\
\end{diagram}
}
\;\; E^x_{\eta \eta'}
\end{equation}
where $n$ is the length of the essential paths $\xi$ and $\xi'$.
\end{proposition}
\begin{proof}
In general we have the following relation between these basis,
\begin{equation}
\label{2.14'}
\xi \otimes \xi' =\sum_{x, \eta, \eta'} B(n , \xi,\xi',x, \eta, \eta') E^x_{\eta \eta'}
\end{equation}
applying $\Phi^y_{\alpha \beta}$ to both sides of (\ref{2.14'}) and using  (\ref{1.4}) and (\ref{2.6})we get,
\begin{equation}
\label{2.15}
B(n , \xi,\xi',x, \eta, \eta') = 
{
\begin{diagram}[size=0.8em,abut]
\,&\xi  &\, \\
\bullet&\rTo~{n} & \bullet  \\
\dTo^{\eta}&  x      &\dTo_{\eta'} \\
\bullet&\rTo & \bullet  \\
\, &\xi'  &\, \\
\end{diagram}
}.
\end{equation} 
\end{proof}
Defining the inverse cells by,
\begin{equation}
\label{2.15'}
\sum_{x,\alpha ,\beta} 
{
\begin{diagram}[size=0.8em,abut]
\,&\xi_1  &\, \\
\bullet&\rTo~{n} & \bullet  \\
\dTo^{\eta}&  x^{-1}      &\dTo_{\eta'} \\
\bullet&\rTo & \bullet  \\
\, &\xi'_1  &\, \\
\end{diagram}
}
{
\begin{diagram}[size=0.8em,abut]
\,&\xi_2  &\, \\
\bullet&\rTo~{n} & \bullet  \\
\dTo^{\eta}&  x      &\dTo_{\eta'} \\
\bullet&\rTo & \bullet  \\
\, &\xi'_2  &\, \\
\end{diagram}
}
=
\delta_{\xi_1 \xi_2}\delta_{\xi'_1 \xi'_2}
\end{equation}
we have,
\begin{equation}
\label{2.15''}
E^x_{\eta \eta'} = \sum_{\xi , \xi'}
{
\begin{diagram}[size=0.8em,abut]
\,&\xi  &\, \\
\bullet&\rTo~{n} & \bullet  \\
\dTo^{\eta}&  x^{-1}      &\dTo_{\eta'} \\
\bullet&\rTo & \bullet  \\
\, &\xi'  &\, \\
\end{diagram}
}
\xi \otimes \xi'
\end{equation}
replacing (\ref{2.13}) in (\ref{2.15''}) the following also holds,
\begin{equation}
\label{2.15'''}
\sum_{\xi ,\xi'} 
{
\begin{diagram}[size=0.8em,abut]
\,&\xi  &\, \\
\bullet&\rTo~{n} & \bullet  \\
\dTo^{\eta}&  x^{-1}      &\dTo_{\eta'} \\
\bullet&\rTo & \bullet  \\
\, &\xi'  &\, \\
\end{diagram}
}
{
\begin{diagram}[size=0.8em,abut]
\,&\xi  &\, \\
\bullet&\rTo~{n} & \bullet  \\
\dTo^{\rho}&  y      &\dTo_{\rho'} \\
\bullet&\rTo & \bullet  \\
\, &\xi'  &\, \\
\end{diagram}
}
=\delta_{xy}
\delta_{\eta \rho}\delta_{\eta' \rho'}
\end{equation}
The result that follows will be very useful for further developments,
\begin{proposition}
For any $(\xi \otimes \xi')_{\mathcal P} \in End^{gr}({\mathcal P})$ there exists a unique element in $End^{gr}({\mathcal E})$ given by,
\begin{equation}
\label{2.16'}
(\xi \otimes \xi')_{\mathcal E} = \sum_{x, \alpha ,\beta}
{
\begin{diagram}[size=0.8em,abut]
\,&\xi  &\, \\
\bullet&\rTo~{n} & \bullet  \\
\dTo^{\alpha}&  x      &\dTo_{\beta} \\
\bullet&\rTo & \bullet  \\
\, &\xi'  &\, \\
\end{diagram}
}
\; 
E^x_{\alpha \beta}
\end{equation}
in such a way that{\footnote{In eq. (\ref{2.16''}) we use the following definition for the concatenation product of elements in $End^{gr}({\mathcal P})$,
\begin{equation}
\label{2.164}
(\xi_1 \otimes \xi'_1)_{\mathcal P} \star (\xi_2 \otimes \xi'_2)_{\mathcal P}
= (\xi_1 \star \xi_2)_{\mathcal P} \otimes (\xi'_1 \star \xi'_2)_{\mathcal P}
\end{equation}}}
,
\begin{eqnarray}
\label{2.16''}
\Phi^x_{\alpha \beta}( (\xi \otimes \xi')_{\mathcal P})&=&
\Phi^x_{\alpha \beta}( (\xi \otimes \xi')_{\mathcal E})
\nonumber\\
\Phi^x_{\alpha \beta}( (\xi_1 \otimes \xi'_1)_{\mathcal E}
 \cdot (\xi_2 \otimes \xi'_2)_{\mathcal E}) &=&
\Phi^x_{\alpha \beta}( (\xi_1 \otimes \xi'_1)_{\mathcal P} \star (\xi_2 \otimes \xi'_2)_{\mathcal P})
\end{eqnarray}
\end{proposition}
\begin{proof}
The first equation in (\ref{2.16''}) is a consequence of the definition of
$(\xi \otimes \xi')_{\mathcal E}$ in (\ref{2.16'}) and the remark, following 
eq.(\ref{1.6'}), about the extension of the definition of connections for elements
in $End^{gr}({\mathcal P})$. For the second equation in (\ref{2.16''}) we have,
\begin{eqnarray}
\label{2.16'''}
\Phi^x_{\alpha \beta}( (\xi_1 \otimes \xi'_1)_{\mathcal E}
 \cdot (\xi_2 \otimes \xi'_2)_{\mathcal E}) &= & 
 \sum_{x_1 , \alpha_1 , \beta_1, x_2 , \alpha_2 , \beta_2}
{
\begin{diagram}[size=0.8em,abut]
\,&\xi_1  &\, \\
\bullet&\rTo~{n} & \bullet  \\
\dTo^{\alpha_1}&  x_1      &\dTo_{\beta_1} \\
\bullet&\rTo & \bullet  \\
\, &\xi'_1  &\, \\
\end{diagram}
}
\;{
\begin{diagram}[size=0.8em,abut]
\,&\xi_2  &\, \\
\bullet&\rTo~{n} & \bullet  \\
\dTo^{\alpha_2}&  x_2      &\dTo_{\beta_2} \\
\bullet&\rTo & \bullet  \\
\, &\xi'_2  &\, \\
\end{diagram}
}
\Phi^x_{\alpha \beta}
(E^{x_1}_{\alpha_1 \beta_1} \cdot E^{x_2}_{\alpha_2 \beta_2})
  \nonumber\\
&= &\sum_{\beta_1}
{
\begin{diagram}[size=0.8em,abut]
\,&\xi_1  &\, \\
\bullet&\rTo~{n} & \bullet  \\
\dTo^{\alpha}&  x      &\dTo_{\beta_1} \\
\bullet&\rTo & \bullet  \\
\, &\xi'_1  &\, \\
\end{diagram}
}
\;{
\begin{diagram}[size=0.8em,abut]
\,&\xi_2  &\, \\
\bullet&\rTo~{n} & \bullet  \\
\dTo^{\beta_1}&  x      &\dTo_{\beta} \\
\bullet&\rTo & \bullet  \\
\, &\xi'_2  &\, \\
\end{diagram}
}
=
\Phi^x_{\alpha \beta}( (\xi_1 \otimes \xi'_1)_{\mathcal P} \star (\xi_2 \otimes \xi'_2)_{\mathcal P})
\end{eqnarray}
where we have employed eq.(\ref{2.1}) and (\ref{2.6}) to write the second equality and (\ref{1.6'}) for the last equality.
\end{proof}
\begin{proposition}
The following holds,
\begin{eqnarray}
\label{2.16}
<E^x_{\eta \eta'}|\xi \otimes \xi'>&= &
{
\begin{diagram}[size=0.8em,abut]
\,&\xi  &\, \\
\bullet&\rTo~{n} & \bullet  \\
\dTo^{\eta}&  x      &\dTo_{\eta'} \\
\bullet&\rTo & \bullet  \\
\, &\xi'  &\, \\
\end{diagram}
}
\\ \label{2.18}
<\rho \otimes \rho'|\xi \otimes \xi'>=
\sum_{x,\eta ,\eta'}<\rho \otimes \rho'|E^x_{\eta \eta'}><E^x_{\eta \eta'}|\xi \otimes \xi'>
& =& 
\sum_{x,\eta ,\eta'}
{\overline{
\begin{diagram}[size=0.8em,abut]
\,&\rho  &\, \\
\bullet&\rTo~{n} & \bullet  \\
\dTo^{\eta}&  x      &\dTo_{\eta'} \\
\bullet&\rTo & \bullet  \\
\, &\rho'  &\, \\
\end{diagram}
}}
{
\begin{diagram}[size=0.8em,abut]
\,&\xi  &\, \\
\bullet&\rTo~{n} & \bullet  \\
\dTo^{\eta}&  x      &\dTo_{\eta'} \\
\bullet&\rTo & \bullet  \\
\, &\xi'  &\, \\
\end{diagram}
}
% \\ \label{2.19}
%{
%\begin{diagram}[size=0.8em,abut]
%\,&\xi_1  &\, \\
%\bullet&\rTo~{n} & \bullet  \\
%\dTo^{\eta}&  x^{-1}      &\dTo_{\eta'} \\
%\bullet&\rTo & \bullet  \\
%\, &\xi'_1  &\, \\
%\end{diagram}
%}
%&=&
%\frac{1}{|\xi \otimes \xi'|^2}
%{\overline{
%\begin{diagram}[size=0.8em,abut]
%\,&\xi  &\, \\
%\bullet&\rTo~{n} & \bullet  \\
%\dTo^{\eta}&  x      &\dTo_{\eta'} \\
%\bullet&\rTo & \bullet  \\
%\, &\xi'  &\, \\
%\end{diagram}
%}}
\end{eqnarray}
\end{proposition}
\begin{proof}
Taking scalar products in both sides of (\ref{2.13}) and using
(\ref{2.4'}) you get the first equality. The other follows from orthogonality and completeness of the 
$|E^x_{\eta \eta'}>$ basis.
\end{proof}
Furthermore as shown in appendix B the form of the scalar product is restricted to,
\begin{proposition}
\label{2.19}
\begin{equation}
\label{}
<\rho \otimes \rho'|\xi \otimes \xi'> \propto \delta_{\rho \xi} \delta_{\rho' \xi'}
\end{equation}
\end{proposition}

%*****************************************************************

\section{Weak bialgebra structure}
\subsection{Product}
The expression of the product in the basis of endomorphisms of essential paths is given by the following.
\begin{proposition}
We have,
\begin{equation}
\label{2.19'}
\xi_1 \otimes \xi'_1 \cdot \xi_2 \otimes \xi'_2 =
\sum_{n_3 , \xi_3 , \xi'_3} P^{\xi_1 \xi_2 \xi_3}_{\xi'_1 \xi'_2 \xi'_3} \;\;\xi_3 \otimes \xi'_3
\end{equation}
where,
\begin{equation}
\label{2.20}
P^{\xi_1 \xi_2 \xi_3}_{\xi'_1 \xi'_2 \xi'_3}=
\sum_{x, \eta_1, \eta_2 ,\eta_3}
{
\begin{diagram}[size=0.8em,abut]
\,&\xi_1  &\, \\
\bullet&\rTo~{n_1} & \bullet  \\
\dTo^{\eta_1}&  x      &\dTo_{\eta_2} \\
\bullet&\rTo & \bullet  \\
\, &\xi'_1  &\, \\
\end{diagram}
}\;
{
\begin{diagram}[size=0.8em,abut]
\,&\xi_2  &\, \\
\bullet&\rTo~{n_2} & \bullet  \\
\dTo^{\eta_2}&  x      &\dTo_{\eta_3} \\
\bullet&\rTo & \bullet  \\
\, &\xi'_2  &\, \\
\end{diagram}
}\;
{
\begin{diagram}[size=0.8em,abut]
\,&\xi_3  &\, \\
\bullet&\rTo~{n} & \bullet  \\
\dTo^{\eta_1}&  x^{-1}      &\dTo_{\eta_3} \\
\bullet&\rTo & \bullet  \\
\, &\xi'_3  &\, \\
\end{diagram}
}
\end{equation}
\end{proposition}
\begin{proof}
Replacing (\ref{2.13}) in the l.h.s. of (\ref{2.19'}), using (\ref{2.1}) and employing (\ref{2.14'}) you get 
the r.h.s. of (\ref{2.19'}) and (\ref{2.20}).
\end{proof}
%%%%%%%%%%%%%%%%%%%%%%%%%%%%%%%%%%%%%%%%%%%%%%%%%%%%%%%%%%%%%%%%%%%%%%%%%%%%%%%%%%%%%%%%%
\subsection{Coproduct}
The definition of a tensor product representation, as given by (\ref{1.7}), can be rephrased in terms of a coproduct.
\begin{proposition}
The following coproduct,
\begin{equation}
\label{2.21}
\Delta(\xi \otimes \xi') = 
\sum_{\xi_i}
(\xi \otimes \xi^i)_{\mathcal E} \otimes(\xi_i \otimes \xi')_{\mathcal E}
\end{equation}
where $\xi_i(\xi^i)$ are elements of the (dual)elementary paths basis of ${\mathcal E}(\hat{\mathcal E})$ respectively. This coproduct implies  (\ref{1.7}),
it is a coassociative algebra morphism and it satisfies,
\begin{equation}
\label{2.22}
\Delta((\xi \otimes \xi')^*)=
\Delta(\xi \otimes \xi')^*
\qquad (a \otimes b)^* = a^* \otimes b^*
\end{equation}
\end{proposition}
\begin{proof}
Replacing eq.(\ref{2.21}) in the r.h.s. of,
\begin{equation}
\label{2.23}
\Phi_{\alpha_1 \circ \alpha_2 ,\beta_1 \circ \beta_2}^{x\otimes y} (\xi \otimes \xi') =
\Phi_{\alpha_1 \beta_1}^{x_1} \otimes \Phi_{\alpha_2 \beta_2}^{x_2}
(\Delta (\xi \otimes \xi'))
\end{equation}
using (\ref{1.4}) and (\ref{2.16''}) you get the same as the r.h.s. of (\ref{1.7}).
Coassociativity is a simple verification.
. In order to verify eq. (\ref{2.22}) 
note that using (\ref{2.13}) and  (\ref{2.4}) 
you obtain,
\begin{equation}
\label{2.25}
(\xi \otimes \xi')^* = 
\sum_{x , \eta , \eta'}
{\overline
{
\begin{diagram}[size=0.8em,abut]
\,&\xi  &\, \\
\bullet&\rTo~{n_2} & \bullet  \\
\dTo^{\eta}&  x      &\dTo_{\eta'} \\
\bullet&\rTo & \bullet  \\
\, &\xi'  &\, \\
\end{diagram}
}}\;
E^x_{\eta' \eta}
\end{equation}
on the other hand using  (\ref{2.13}) and reflection you have.
\begin{equation}
\label{2.25'}
{\tilde \xi} \otimes {\tilde \xi'} = 
\sum_{x , \eta' , \eta}
{
\begin{diagram}[size=0.8em,abut]
\,&{\tilde \xi}  &\, \\
\bullet&\rTo~{n_2} & \bullet  \\
\dTo^{\eta'}&  x      &\dTo_{\eta} \\
\bullet&\rTo & \bullet  \\
\, &{\tilde \xi}'  &\, \\
\end{diagram}
}\,
E^x_{\eta' \eta}
=
\sum_{x , \eta' , \eta}
\sqrt{\frac{\mu_i^{\xi} \mu_f^{\xi'}}{\mu_f^{\xi} \mu_i^{\xi'} }
}\;\;
{\overline{
\begin{diagram}[size=0.8em,abut]
\,&\xi  &\, \\
\bullet&\rTo~{n_2} & \bullet  \\
\dTo^{\eta}&  x      &\dTo_{\eta'} \\
\bullet&\rTo & \bullet  \\
\, & \xi'  &\, \\
\end{diagram}
}
} \, E^x_{\eta' \eta}
=
\sqrt{\frac{\mu_i^{\xi} \mu_f^{\xi'}}{\mu_f^{\xi} \mu_i^{\xi'} }
}\,
(\xi \otimes \xi')^*
\end{equation}
where in the last equality we have used (\ref{2.25}).
Using (\ref{2.25'}) it is simple to verify  (\ref{2.22}).
The morphism property of the coproduct is dealt with in appendix C.
\end{proof}
The coproduct can be expressed in the basis $\{E^x_{\alpha \beta}\}$
as stated below.
\begin{proposition}
We have,
\begin{equation}
\label{2.26}
\Delta(E^x_{\eta \eta'})=
\sum_{
x_1,  \eta_1 , \eta'_1 
x_2,  \eta_2 ,\eta'_2 
}
{\tilde P}^{x\;\eta_1 \eta'_1 \,\eta_2 \eta'_2 }_{\eta \eta' x_1  \,\; x_2 }
E^{x_1}_{\eta_1 \eta'_1} \otimes E^{x_2}_{\eta_2 \eta'_2}
\end{equation}
where,
\begin{equation}
\label{2.27}
{\tilde P}^{x\;\eta_1 \eta'_1 \,\eta_2 \eta'_2 }_{\eta \eta' x_1  \,\; x_2 }
=
\sum_{n,\xi_1 , \xi_2 ,\xi_3} 
\;
{
\begin{diagram}[size=0.8em,abut]
\,&\xi_1  &\, \\
\bullet&\rTo~{n_1} & \bullet  \\
\dTo^{\eta}&  x^{-1}      &\dTo_{\eta'} \\
\bullet&\rTo & \bullet  \\
\, &\xi_3  &\, \\
\end{diagram}
}
 \;
{
\begin{diagram}[size=0.8em,abut]
\,&\xi_1  &\, \\
\bullet&\rTo~{n_1} & \bullet  \\
\dTo^{\eta_1}&  x_1      &\dTo_{\eta'_1} \\
\bullet&\rTo & \bullet  \\
\, &\xi_2  &\, \\
\end{diagram}
}
{
\begin{diagram}[size=0.8em,abut]
\,&\xi_2  &\, \\
\bullet&\rTo~{n_1} & \bullet  \\
\dTo^{\eta_2}&  x_2      &\dTo_{\eta'_2} \\
\bullet&\rTo & \bullet  \\
\, &\xi_3  &\, \\
\end{diagram}
}
\end{equation}
\end{proposition}
\begin{proof}
Replacing (\ref{2.15''}) in the l.h.s. of (\ref{2.26}), employing (\ref{2.21})
and using (\ref{2.13}) two times you get the result.
\end{proof}
%%%%%%%%%%%%%%%%%%%%%%%%%%%%%%%%%%%%%%%%%%%%%%%%%%%%%%%%%%%%%%%%%%
\subsection{Counit}
Regarding the counit the following holds.
\begin{proposition}
A counit satisfying the property,
\begin{equation}
\label{2.28}
(\epsilon \otimes \one) \Delta  =\one
= (\one \otimes \epsilon  ) \Delta 
\end{equation}
is given by,
\begin{equation}
\label{2.29}
\epsilon(\xi \otimes \xi') = \delta_{\xi\xi'} 
\end{equation}
\end{proposition}
\begin{proof}
Just replace (\ref{2.29}) in (\ref{2.28}).
\end{proof}
It is important to note that the relation $\Delta(\one) = \one \otimes \one$ does not hold in the DTA, which is therefore {\em{not}} a Hopf algebra.
The structure of the DTA appearing up to this part corresponds to a 
weak bialgebra in the terminology of  \cite{nill} definition 2.1. As we shall see in the next section there is also an antipode and the DTA is actually a weak Hopf algebra or  "quantum groupo\"\i d" (this general notion is discussed in   \cite{bs, NikVainFiniteQG, NVdepth, NikshychVainerman, nill}).
%%%%%%%%%%%%%%%%%%%%%%ANTIPODE%%%%%%%%%%%%%%%%%%%%%%%%%%%%%
\section{Weak *-Hopf algebra structure. Antipode}

According to Theorem 8.7 of \cite{nill}, we have

\begin{theorem}[Nill]
Let $(A,\one,\Delta,\epsilon)$ be a weak bialgebra and $S: A \to A$
be a bialgebra antiautomorphism. If there exists a non-degenerate 
linear functional $\lambda: A \to {\mathbb C}$ such that
{\footnote{In eq.(\ref{2.31}) we employed Sweedler's notation for the coproduct, i.e.,
\begin{equation}
\label{2.30}
\Delta (a ) = a_{(1)} \otimes a_{(2)} \qquad
\Delta (b ) = b_{(1)} \otimes b_{(2)}
\end{equation} }},
\begin{equation}
\label{2.31}
a_{(1)} \lambda ( b a_{(2)}) = S (b_{(1)}) \lambda(b_{(2)}a) \qquad, \forall a,b \;\in A
\end{equation} 
then  $S$ is an antipode and $A$ is a weak Hopf algebra. 
\end{theorem}
Below we prove the following result,
\begin{theorem}
\label{wh}
The DTA algebra has the structure of a weak $*$-Hopf algebra
with product given by (\ref{2.1}), unit given by (\ref{2.2}),
coproduct (\ref{2.21}), counit (\ref{2.29}) , star (\ref{2.4}) and antipode,
\begin{equation}
\label{2.32}
S({ \xi}_i \otimes { \xi}^j) = {\tilde{\xi}_j} \otimes {\tilde{\xi}^i}
\end{equation}
\end{theorem}
\begin{proof}
Take,
\begin{equation}
\label{2.41}
\lambda(E^{x}_{\eta {\eta}'})= \delta_{x {\mathbf 0}}\delta_{\eta {\eta}'}
\end{equation}
where $\mathbf 0$ denotes the trivial representation as described in appendix D. Let,
\begin{equation}
\label{2.42}
a= \xi_1 \otimes {\xi'}_1  \qquad , 
b= \xi_2 \otimes {\xi'}_2
\end{equation}
in (\ref{2.31}). Using (\ref{2.21}) we obtain,
\begin{equation}
\label{2.43}
\Delta(a) = 
\sum_{\rho_1}
(\xi_1 \otimes \rho_1)\otimes (\rho_1 \otimes \xi'_1)
\qquad
\Delta(b) = 
\sum_{\rho_2}
(\xi_2 \otimes \rho_2)\otimes (\rho_1 \otimes \xi'_1)
\end{equation}
therefore consider,
\begin{eqnarray}
\label{2.44}
A&=&\sum_{\rho_1} 
(\xi_1 \otimes \rho_1) 
\;\lambda( (\xi_2 \otimes \xi'_2 ) \cdot
(\rho_1 \otimes \xi'_1) )- \sum_{\rho_2}
S( (\xi_2 \otimes \rho_2))
\;\lambda( (\rho_2 \otimes \xi'_2) \cdot 
(\xi_1 \otimes \xi'_1 )  ) 
\end{eqnarray}
so that eq.(\ref{2.31}) becomes $A=0$. Now in order to evaluate the products in the arguments of $\lambda$ we 
rewrite (\ref{2.44}) in terms of the basis $\{ E^x_{\eta \eta'} \}$
to obtain,
\begin{eqnarray}
\label{2.45}
A&=&\sum_{ x_1, \eta_1 ,\eta'_1 , x_2, \eta_2 ,\eta'_2}
\left[\sum_{\rho_1} 
(\xi_1 \otimes \rho_1 )
{
\begin{diagram}[size=0.8em,abut]
\,&\xi_2  &\, \\
\bullet&\rTo~{n} & \bullet  \\
\dTo^{\eta_1}&  x_1      &\dTo_{\eta'_1} \\
\bullet&\rTo & \bullet  \\
\, & \xi'_2  &\, \\
\end{diagram}
}
\;\;
{
\begin{diagram}[size=0.8em,abut]
\,&\rho_1  &\, \\
\bullet&\rTo~{n} & \bullet  \\
\dTo^{\eta_2}&  x_2      &\dTo_{\eta'_2} \\
\bullet&\rTo & \bullet  \\
\, & \xi'_1  &\, \\
\end{diagram}
}\right.
\nonumber\\ 
&& \left. \;\; -\sum_{\rho_2}
S( (\xi_2 \otimes \rho_2))\;\;
{
\begin{diagram}[size=0.8em,abut]
\,&\rho_2  &\, \\
\bullet&\rTo~{n} & \bullet  \\
\dTo^{\eta_1}&  x_1      &\dTo_{\eta'_1} \\
\bullet&\rTo & \bullet  \\
\, & \xi'_2  &\, \\
\end{diagram}
}
\;\;
{
\begin{diagram}[size=0.8em,abut]
\,&\xi_1  &\, \\
\bullet&\rTo~{n} & \bullet  \\
\dTo^{\eta_2}&  x_2      &\dTo_{\eta'_2} \\
\bullet&\rTo & \bullet  \\
\, & \xi'_1  &\, \\
\end{diagram}
}
\right] 
\;\lambda(E^x_{\eta_1 \eta'_1} \cdot E^x_{\eta_2 \eta'_2})
 \end{eqnarray}
now we use (\ref{2.1}) and  (\ref{2.41}) to obtain
$\lambda(E^{x_2}_{\eta_2 \eta'_2}\cdot E^{x_1}_{\eta_1 \eta'_1} )
=\delta_{x_1 {\mathbf 0}}\delta_{x_1 x_2}\delta_{\eta'_1 \eta_2}\delta_{\eta_1 \eta'_2}$, replacing in (\ref{2.45}) and using reflection we obtain,
\begin{eqnarray}
\label{2.46}
A&=&\sum_{  \eta_1 , \eta_2 }
\left[ \sum_{\rho_1} 
\sqrt{\frac{\mu_f^{\xi_2} \mu_i^{\xi'_2}}{\mu_i^{\xi_2} \mu_f^{\xi'_2} }
}
(\xi_1 \otimes \rho_1 )
{\overline{
\begin{diagram}[size=0.8em,abut]
\,&{\tilde \xi}_2  &\, \\
\bullet&\rTo~{n} & \bullet  \\
\dTo^{\eta_1}&  {\mathbf 0}      &\dTo_{\eta_2} \\
\bullet&\rTo & \bullet  \\
\, & {\tilde \xi}'_2  &\, \\
\end{diagram}
}}
\;\;
{
\begin{diagram}[size=0.8em,abut]
\,&\rho_1  &\, \\
\bullet&\rTo~{n} & \bullet  \\
\dTo^{\eta_1}&  {\mathbf 0}      &\dTo_{\eta_2} \\
\bullet&\rTo & \bullet  \\
\, & \xi'_1  &\, \\
\end{diagram}
}\right.\nonumber\\
 \;\;&&\left. -
\sqrt{\frac{\mu_f^{\rho_2} \mu_i^{\xi'_2}}{\mu_i^{\rho_2} \mu_f^{\xi'_2} }
}
 S( (\xi_2 \otimes \rho_2))\;\;          %%%%%%%%%%%%%%%%%%%%%%%
{\overline{
\begin{diagram}[size=0.8em,abut]
\,&{\tilde \rho_2}  &\, \\
\bullet&\rTo~{n} & \bullet  \\
\dTo^{\eta_1}&  {\mathbf 0}      &\dTo_{\eta_2} \\
\bullet&\rTo & \bullet  \\
\, & {\tilde \xi}'_2  &\, \\
\end{diagram}
}}
\;\;
{
\begin{diagram}[size=0.8em,abut]
\,&\xi_1  &\, \\
\bullet&\rTo~{n} & \bullet  \\
\dTo^{\eta_1}&  {\mathbf 0}      &\dTo_{\eta_2} \\
\bullet&\rTo & \bullet  \\
\, & \xi'_1  &\, \\
\end{diagram}
}
\right]\nonumber\\ 
 %\end{eqnarray}
&&
=\sqrt{\frac{\mu_f^{\xi_2} \mu_i^{\xi'_2}}{\mu_i^{\xi_2} \mu_f^{\xi'_2}}}
\sum_{\rho_1} (\xi_1 \otimes \rho_1 )
<{\tilde \xi}_2 \otimes {\tilde \xi}'_2|\rho_1 \otimes \xi'_1 >_0
\nonumber\\ 
&& - \sum_{\rho_2 }
\sqrt{\frac{\mu_f^{\rho_2} \mu_i^{\xi'_2}}{\mu_i^{\rho_2} \mu_f^{\xi'_2} }}
S( (\xi_2 \otimes \rho_2))\;\;         
<{\tilde \rho}_2\otimes {\tilde \xi}'_2 |\xi_1 \otimes \xi'_1 >_0
\end{eqnarray}
where we have employed (\ref{2.19}) and the definition of scalar product in a representation appearing in Appendix B. Using the results in appendix B and noting, as shown in appendix D, that in the representation $\mathbf 0$ the only non-vanishing cells are those with equal upper and lower horizontal paths we  conclude that there is the following factorization in eq.(\ref{2.46}),
\begin{equation}
\label{2.47}
A=\sqrt{\frac{\mu_f^{\xi_2} \mu_i^{\xi'_2}}{\mu_i^{\xi_2} \mu_f^{\xi'_2}}}
<\xi'_1 \otimes \xi'_1|\xi'_1 \otimes \xi'_1>
(\xi_1 \otimes {\tilde \xi}_2 \;\;\delta_{\xi'_1 {\tilde \xi}'_2}
-
S( (\xi_2 \otimes {\tilde \xi}_1))\;\;\delta_{\xi'_1 {\tilde \xi}'_2}
)
\end{equation}
hence $A=0$ implies,
\begin{equation}
\label{2.48}
S(\xi \otimes \xi') = {\tilde \xi}' \otimes {\tilde \xi}
\end{equation}
Next we show $S$ is a bialgebra antiautomorphism, i.e., 
\begin{equation}
\label{2.33}
S(a \cdot b) = S(b) S(a)
\end{equation}
In order to do this we first express the antipode in the $\{ E^x_{\eta \eta'} \}$ basis. Replacing (\ref{2.13}) in (\ref{2.48})
we obtain,
\begin{equation}
\label{2.34}
\sum_{x, \eta, \eta'}
{
\begin{diagram}[size=0.8em,abut]
\,&\xi  &\, \\
\bullet&\rTo~{n} & \bullet  \\
\dTo^{\eta}&  x      &\dTo_{\eta'} \\
\bullet&\rTo & \bullet  \\
\, &\xi'  &\, \\
\end{diagram}
}
\;\; S ( E^x_{\eta \eta'})
=
\sum_{y, \rho, \rho'}
{
\begin{diagram}[size=0.8em,abut]
\,&{\tilde \xi}'  &\, \\
\bullet&\rTo~{n} & \bullet  \\
\dTo^{\rho}&  x      &\dTo_{\rho'} \\
\bullet&\rTo & \bullet  \\
\, &{\tilde \xi}  &\, \\
\end{diagram}
}
\;\; E^y_{\rho \rho'}
\end{equation}
multiplying this equation by,
\begin{equation}
\label{2.35}
{{
\begin{diagram}[size=0.8em,abut]
\,&\xi  &\, \\
\bullet&\rTo~{n} & \bullet  \\
\dTo^{\lambda}&  z^{-1}      &\dTo_{\lambda'} \\
\bullet&\rTo & \bullet  \\
\, &{\tilde \xi'}  &\, \\
\end{diagram}
}}
\end{equation}
summing up over $\xi$ and $\xi'$ and using (\ref{2.15'''}) we have,
\begin{equation}
\label{2.36}
S(E^z_{\lambda \lambda'})=
\sum_{y, \rho, \rho'}
\sum_{\xi \xi'} 
{{
\begin{diagram}[size=0.8em,abut]
\,&\xi  &\, \\
\bullet&\rTo~{n} & \bullet  \\
\dTo^{\lambda}&  z^{-1}      &\dTo_{\lambda'} \\
\bullet&\rTo & \bullet  \\
\, &{\tilde \xi'}  &\, \\
\end{diagram}
}}\;\;
{
\begin{diagram}[size=0.8em,abut]
\,&{\tilde \xi}'  &\, \\
\bullet&\rTo~{n} & \bullet  \\
\dTo^{\rho}&  x      &\dTo_{\rho'} \\
\bullet&\rTo & \bullet  \\
\, &{\tilde \xi}  &\, \\
\end{diagram}
}
E^y_{\rho \rho'}
\end{equation}
now using two times reflection we have that,
\begin{equation}
\label{2.37}
{
\begin{diagram}[size=0.8em,abut]
\,&{\tilde \xi}'  &\, \\
\bullet&\rTo~{n} & \bullet  \\
\dTo^{\rho}&  x      &\dTo_{\rho'} \\
\bullet&\rTo & \bullet  \\
\, &{\tilde \xi}  &\, \\
\end{diagram}
}
=
{
\begin{diagram}[size=0.8em,abut]
\,&\xi  &\, \\
\bullet&\rTo~{n} & \bullet  \\
\dTo^{{\tilde \rho}'}&  x      &\dTo_{{\tilde \rho}} \\
\bullet&\rTo & \bullet  \\
\, &\xi'  &\, \\
\end{diagram}
}
\end{equation}
replacing in (\ref{2.36}) and using once more (\ref{2.15'''})
we obtain,
\begin{equation}
\label{2.38}
S(E^z_{\lambda \lambda'})=
E^z_{{\tilde \lambda}' {\tilde \lambda}}
\end{equation}
hence,
\begin{equation}
\label{2.39}
S(E^{x_1}_{\eta_1 {\eta_1}'} \cdot E^{x_2}_{\eta_2 {\eta_2}'})=
\delta_{{\eta_1}' \eta_2} \delta_{x_1 x_2}
S(E^{x_1}_{\eta_1 {\eta_2}'})=
\delta_{{\eta_1}' \eta_2} \delta_{x_1 x_2}
E^{x_1}_{{{\tilde \eta}_2}' {\tilde \eta}_1 }
\end{equation}
on the other hand,
\begin{equation}
\label{2.40}
S(E^{x_1}_{\eta_1 {\eta_1}'}) \cdot S(E^{x_2}_{\eta_2 {\eta_2}'})=
E^{x_2}_{{{\tilde \eta}_2}' {\tilde \eta}_2 } \cdot
E^{x_1}_{{{\tilde \eta}_1}' {\tilde \eta}_1 }
=\delta_{{\eta_1}' \eta_2} \delta_{x_1 x_2}
E^{x_1}_{{{\tilde \eta}_2}' {\tilde \eta}_1 }
\end{equation}
that proves (\ref{2.33}).
\end{proof}

%%%%%%%%%%%%%%%%%%%%EXAMPLE%%%%%%%%%%%%%%%%%%%%%%%%%%%%%%%%%%

\begin{example}[The case of $A_3$]
For the case of the fundamental representation, that we denote in this case by $\mathbf 1$, using 
(\ref{2.13}) we can express elements of $End^{gr}(\mathcal E)$
in terms of matrix units. The connections involved can be calculated from the basic ones and horizontal concatenation of cells. We obtain,
\begin{equation}
\label{}
\begin{array}{llll}
0 \otimes 1=  E^{\mathbf 1}_{(01)(01)}\;,\;
&
1 \otimes 0=  E^{\mathbf 1}_{(10)(10)}\;,\;
&
1 \otimes 2=  E^{\mathbf 1}_{(12)(12)}\;,\;
&
2 \otimes 1=  E^{\mathbf 1}_{(21)(21)}\\
01 \otimes 12=  E^{\mathbf 1}_{(01)(12)}\;,\;  
&
01 \otimes 10=  E^{\mathbf 1}_{(01)(10)}  \;,\;
&
10 \otimes 01=  E^{\mathbf 1}_{(10)(01)} / \sqrt{2} \;,\;
&
10 \otimes 21=  E^{\mathbf 1}_{(12)(01)} / \sqrt{2} \\
12 \otimes 01=  E^{\mathbf 1}_{(10)(21)} / \sqrt{2} \;,\;
&
12 \otimes 21=  -E^{\mathbf 1}_{(12)(21)} / \sqrt{2} \;,\;
&
21 \otimes 12=  -E^{\mathbf 1}_{(21)(12)} \;,\;
&
21 \otimes 10=  E^{\mathbf 1}_{(21)(10)}  \\
012 \otimes \gamma=  -E^{\mathbf 1}_{(01)(21)} \;,\;
&
210 \otimes \gamma=    -E^{\mathbf 1}_{(21)(01)} \;,\;
&
\gamma \otimes 210=    -E^{\mathbf 1}_{(12)(10)} \;,\;
&
\gamma \otimes 012=    -E^{\mathbf 1}_{(10)(12)} \\
\end{array}
\end{equation}
This can be summarized in the following matrix,
\begin{equation}
\label{2.41'}
\begin{array}{l}
01\\
10\\
12\\
21\\
\end{array}
\left(
\begin{array}{llll}
01&r_0 l_1&r_0 r_1&-d \gamma \\
l_1 r_0/\sqrt{2}&10&-\gamma d&r_1 r_0 /\sqrt{2} \\
l_1 l_2/\sqrt{2}&-\gamma g&12&-r_1 l_2/\sqrt{2} \\
\gamma &l_2 l_1&-l_2 r_1&21 \\
\end{array}
\right)
\end{equation}
that should be interpreted in the following way. The matrix representing the element $\xi_i \otimes \xi^j$ in the representation $f$ is the one obtained from (\ref{2.41'}) by replacing in it $\xi_i \, \xi^j$ by 
$1$ and all the others by $0$.
Now we consider the tensor product representation 
${\mathbf 1} \otimes {\mathbf 1}$. Taking into account (\ref{1.7}) and with the same conventions as in (\ref{2.41}) you obtain the following matrix,
\begin{equation}
\label{2.42'}
\begin{array}{l}
010\\
212\\
101\\
121\\
012\\
210\\
\end{array}
\left(
\begin{array}{llllll}
00&dd&1/\sqrt{2}r_0 r_0&1/\sqrt{2}r_0 r_0&0&0  \\
gg&22&1/\sqrt{2}l_2 l_2&1/\sqrt{2}l_2 l_2&0&0 \\
1/\sqrt{2}l_1 l_1&1/\sqrt{2}r_1 r_1&11&\gamma \gamma&
1/\sqrt{2}l_1 r_1 &1/\sqrt{2}r_1 l_1\\
1/\sqrt{2}l_1 l_1&1/\sqrt{2}r_1 r_1&\gamma \gamma&11&
-1/\sqrt{2}l_1 r_1 &1/\sqrt{2}r_1 l_1\\
0&0&1/\sqrt{2} r_0 l_2&-1/\sqrt{2} r_0 l_2&02&dg\\
0&0&1/\sqrt{2} l_2 r_0&-1/\sqrt{2}  l_2 r_0&gd&20\\
\end{array}
\right)
\end{equation}
Note that the labels of files and columns are two step paths
(not in general essential) in the graph $A_3$.
In order to decompose this tensor product representation in direct sum of irreps it is worth noting that,
\begin{enumerate}
\item In many instances of elements in $End^{gr}(\mathcal E)$
say $\xi_i \otimes \xi^j$, the indices corresponding to non-vanishing entries in the 
irrep under study are completely determined because there is only one possibility for them.
For example take the element $0 \otimes 0$ then the only possibility
for the step two vertical paths is $\alpha= (010)$  and $\beta= (010)$
That is the only possible non-zero cell is,
\begin{equation}
\label{2.43'}
{
\begin{diagram}[size=0.8em,abut]
0&   \,&0 \\
\bullet&\rTo & \bullet  \\
\dTo^{(010)}&  f      &\dTo_{(010)} \\
\bullet&\rTo & \bullet  \\
0 &\,  & 0\\
\end{diagram}
}=1
\end{equation}
In other words, the end points of the horizontal essential paths give only one possibility for the vertical (not necessarily essential) paths.
\item Note that in the case of $A_3$ if the two initial or final  points of the horizontal paths are $1$ then in general the associated vertical paths are not fixed and they can be a linear combination of $(101)$ and $(121)$. This combination should be determined in such a way that one gets a block decomposition of the matrix (\ref{2.42'}).
\end{enumerate}
From the above remarks one reduces in this case the block diagonalization problem to the one of $2 \times 2$ matrix.
It is a simple calculation to see that the right vertical labels are the following combinations of $(101)$ and $(121)$,
\begin{equation}
\gamma= 1/\sqrt{2} ( (121) - (101)) \qquad, \; 
\gamma'= 1/\sqrt{2} ( (121) + (101))
\label{2.44'}
\end{equation}
with these horizontal labels one gets for the matrix (\ref{2.42'})
the following,
\begin{equation}
\label{2.45'}
\begin{array}{l}
010 \\
212\\
\gamma'\\
\gamma\\
012\\
210\\
\end{array}
\left(
\begin{array}{llllll}
00&dd&r_0 r_0&0&0&0  \\
gg&22&l_2 l_2&0&0&0 \\
l_1 l_1&r_1 r_1&11+\gamma \gamma&0&0&0\\
0&0&0&11-\gamma \gamma&-l_1 r_1&-r_1 l_1  \\
0&0&0&-r_0 l_2 &02 &dg\\
0&0&0&-l_2 r_0 &gd& 20\\
\end{array}
\right)
\end{equation}
the first block in (\ref{2.45'}) corresponds to the trivial representation $\mathbf 0$ that is described in appendix D. The other block is denoted by $\mathbf 2$ and the following table of tensor product representations decomposition in irreps is easily obtained by employing analogous methods as in the case of ${\mathbf 1} \otimes {\mathbf 1}$,
\begin{eqnarray}
\label{2.46'}
{\mathbf 0}\otimes {\mathbf 1}&=& {\mathbf 1} \nonumber\\
{\mathbf 0}\otimes {\mathbf 2}&=& {\mathbf 2}  \nonumber\\
{\mathbf 1}\otimes {\mathbf 2}&=& {\mathbf 1} \nonumber\\
{\mathbf 1}\otimes {\mathbf 1}&=& {\mathbf 0} \oplus {\mathbf 2} \nonumber\\
\end{eqnarray}
this table itself can be represented by a graph. In this graph each vertex corresponds to an irrep of the DTA of $A_3$. If a line joins two vertices, it means that one irrep can be obtained from the another one by taking tensor product with the fundamental $\mathbf 1$. Notice that the trivial representation $\mathbf 0$ is not 1-dimensional (this is a usual  feature of weak Hopf algebras).
In this case this graph coincides with the graph $A_3$. In the literature this graph is referred to as Ocneanu's graph of quantum symmetries.

 In general, the fusion algebra --that describes the tensor product of representations for the composition product in the dual-- has no reason to be  isomorphic with the algebra of quantum symmetries -- that describes the tensor product of representations for the convolution product  $\cdot$.  However, in the case of $A_n$ graphs (and only in such cases), these two algebras are isomorphic. For these particular Dynkin diagrams, the block decomposition of the  two products (the convolution product $\cdot$ in the DTA and the composition of endomorphisms in its dual)  have the same number of direct summands and these summands have also the same dimensions.

\end{example}

%%%%%%%%%%%%%%%%%%%%%%%%%%%%%%%%%%%%%%%%%%%%%%%%%%%%%%%%%%%%%%%%%
\section*{Appendix A. General form of ADE connections}
\begin{proposition}
Given a cell system where the four sides of the cells correspond to length-one paths 
on the same ADE graph $G$, the following two ``basic'' connections{\footnote{In the drawing below we omit the path indices.}}, \footnote{These expressions were given by A. Ocneanu in various seminars (unpublished) }
(we only give one of the two, but the other is obtained by taking the complex conjugate of the first) satisfy unitarity and reflection,
\begin{equation}
\label{a1}
{
\begin{diagram}[size=0.8em,abut]
v_i&\,&v_l \\
\bullet&\rTo & \bullet    \\
\dTo& f      &\dTo \\
\bullet&\rTo & \bullet   \\
v_k &\, &v_j \\
\end{diagram}
}
=
\delta_{v_k v_l} \, \epsilon + \delta_{v_i v_j} \sqrt{\frac{\mu_{v_k} \mu_{v_l}}{\mu_{v_i} \mu_{v_j}}}
{\bar \epsilon}
\end{equation}
where $\epsilon = i e^{i \pi / 2 N}$ with $N$ the Coxeter number of $G$.
\end{proposition}
\begin{proof}
Reflection is a simple check.  The unitarity conditions is,
\begin{equation}
\label{a2}
A=\sum_{v_l}
{
\begin{diagram}[size=0.8em,abut]
v_i&\,&v_l \\
\bullet&\rTo & \bullet   \\
\dTo&   f     &\dTo \\
\bullet&\rTo & \bullet   \\
v_k &\, &v_j \\
\end{diagram}
}\;\;
{\overline{
\begin{diagram}[size=0.8em,abut]
v_i &\,&v_l \\
\bullet&\rTo & \bullet   \\
\dTo&   f     &\dTo \\
\bullet&\rTo & \bullet   \\
v_m &\, &v_j \\
\end{diagram}
}} = \delta_{v_k v_m}
\end{equation}
Consider now the case when $v_i \neq v_j$. In this case the only term of (\ref{a1}) that contributes is the first
and replacing in (\ref{a2}) you get,
\begin{equation}
\label{a3}
A= \sum_{v_l} \delta_{v_k v_l}\, \epsilon \; \delta_{v_m v_l}\, {\bar \epsilon} = \delta_{v_k v_m}
\end{equation}
as claimed. In the other case $v_i=v_j$ you have,
\begin{equation}
\label{a4}
A= \sum_l 
(\sqrt{\frac{\mu_{v_k} \mu_{v_l}}{\mu_{v_i}^2}} {\bar \epsilon} + \delta_{v_k v_l} \epsilon)
(\sqrt{\frac{\mu_{v_m} \mu_{v_l}}{\mu_{v_i}^2}} \epsilon + \delta_{v_k v_l} {\bar \epsilon})
= \delta_{v_k v_m} + \sqrt{\frac{\mu_{v_m} \mu_{v_l}}{\mu_{v_i}^2}}
[\epsilon^2 + {\bar \epsilon}^2 + \frac{1}{\mu_{v_i}} \sum_{<v_l v_i>} \mu_{v_l}]
\end{equation}
where $<v_l v_i>$ in (\ref{a4}) indicates that the summation is over the vertices $v_l$ in $G$ that are connected with 
the vertex $v_i$. This summation can therefore be expressed in terms of the adjacency matrix $M$ of $G$ and its Perron-Frobenius eigenvector $v_{pf}$ as follows,
\begin{equation}
\label{a5}
\sum_{<v_l v_i>} \mu_{v_l} = 
<row\;v_i\; of\; M ,v_{pf}> 
=<{\hat e}_{v_i} M,v_{pf}>
\end{equation}
where $\{{\hat e}_{v_i}\}$ is a basis vector associated to the vertex $v_i$ of $G$. The scalar product $<,>$
in this vector space ${\mathbb C}^{N_v}$ (where $N_v$ is the number of vertices in $G$)
that appears in (\ref{a5}) is the Euclidean one in this basis and the matrix elements $M_{v_i v_j}$ of the adjacency matrix $M$ vanish unless vertex $v_i$ is connected to vertex $v_j$. 
Now since $M$ is hermitian we have,
\begin{equation}
\label{a6}
\sum_{<v_l v_i>} \mu_{v_l} = 
<{\hat e}_{v_i} M,v_{pf}>=
<{\hat e}_{v_i} ,M v_{pf}>= \beta \, \mu_{v_i}
\end{equation} 
replacing (\ref{a6}) in (\ref{a4}) and noting that for ADE graphs 
\begin{equation}
\label{a7}
\beta = 2 \cos{\pi/N}
\end{equation} 
the second term in (\ref{a4}) cancels since $\epsilon^2 + {\bar \epsilon}^2 + \beta=0$ leading to (\ref{a2}). 
\end{proof}
Furthermore, regarding the uniqueness of connections we have the following result whose proof is a simple check.
\begin{proposition}[Gauge freedom]	
If,
\begin{equation}
\label{a8}
{
\begin{diagram}[size=0.8em,abut]
v_i &\,&v_l \\
\bullet&\rTo & \bullet  \\
\dTo& f      &\dTo \\
\bullet&\rTo & \bullet   \\
v_k & \, &v_j \\
\end{diagram}
}
\end{equation}
satisfies unitarity and reflection then,
\begin{equation}
\label{a9}
{
\begin{diagram}[size=0.8em,abut]
v_i &\,&v_l \\
\bullet&\rTo & \bullet  \\
\dTo& f'      &\dTo \\
\bullet&\rTo & \bullet   \\
v_k &\, &v_j \\
\end{diagram}
}
=
e^{i (\alpha_{v_i} +\alpha_{v_j}-\alpha_{v_k}-\alpha_{v_l})}
{
\begin{diagram}[size=0.8em,abut]
v_i &\,&v_l \\
\bullet&\rTo & \bullet \\
\dTo& f      &\dTo \\
\bullet&\rTo & \bullet  \\
v_k &\, &v_j \\
\end{diagram}
}
\end{equation}
also satisfies them with $\alpha_{v_i} ,\alpha_{v_j},\alpha_{v_k},\alpha_{v_l}$
real numbers associated to the vertices of $G$.
\end{proposition}
\begin{example}[The case of $A_3$]
From example (\ref{ex1}) we know that $\beta=\sqrt{2}$ for $A_3$
hence taking into account (\ref{a7}) we have $N=4$ for this case.
Thus $\epsilon = e^{i 5\pi/8}$. Using (\ref{a1}) and making a gauge transformation as in (\ref{a9}) with
$\alpha_0= \frac{15}{16} \pi \;\,, \alpha_1 =0\;\,,
\alpha_2= \frac{7}{16} \pi$ you get the results appearing in (\ref{1.9}).
\end{example}

%%%%%%%%%%%%%%%%%%%%%%%%%%%%%%%%%%%%%%%%%%%%%%%%%%%%%%%%%%%%%%%%%%
\section*{Appendix B. The scalar product.}
We first remark that the scalar product (\ref{2.4'}) can be written as,
\begin{equation}
\label{d1}
< E^x_{\alpha \beta} | E^y_{\gamma \delta}> = Tr[(E^x_{\alpha \beta})^*  E^y_{\gamma \delta} ]
\end{equation}
this can  easily be verified  by using  (\ref{2.4}), (\ref{2.1}) and (\ref{2.5}).
Using (\ref{d1}) we can write the scalar product (\ref{2.18}) as,
\begin{equation}
\label{dB2}
<\rho \otimes \rho'|\xi \otimes \xi'>= \sum_{x \;irrep.} 
<\rho \otimes \rho'|\xi \otimes \xi'>_x
\end{equation}
where in general for a representation $R$ irreducible or not,
\begin{equation}
\label{dB3}
<\rho \otimes \rho'|\xi \otimes \xi'>_R =
Tr_R [(\Phi^R(\rho \otimes \rho'))^{\dagger} \Phi^R (\xi \otimes \xi')] 
\end{equation}
In order to prove proposition \ref{p2.3} we first show that,
\begin{proposition}
\label{d}
\begin{equation}
\label{dB4}
<\rho \otimes \rho'|\xi \otimes \xi'>_R \propto 
\delta_{\rho \xi} \delta_{\rho' \xi'}
\end{equation}
with $R= f_1 \otimes\cdots \otimes f_n$ for any $n$, where $f_1, \cdots, f_n$ can be any of the fundamentals.
\end{proposition}
\begin{proof}
We will use induction.
For $R=f$ (\ref{dB4}) holds. This can be seen by recalling the expression of the scalar product
(\ref{dB3}) in terms of cells,  
\begin{equation}
\label{dB5}
<\rho \otimes \rho'|\xi \otimes \xi'>_f =
\sum_{\eta ,\eta'}
{\overline{
\begin{diagram}[size=0.8em,abut]
\,&\rho  &\, \\
\bullet&\rTo~{n} & \bullet  \\
\dTo^{\eta}&  f      &\dTo_{\eta'} \\
\bullet&\rTo & \bullet  \\
\, &\rho'  &\, \\
\end{diagram}
}}
{
\begin{diagram}[size=0.8em,abut]
\,&\xi  &\, \\
\bullet&\rTo~{n} & \bullet  \\
\dTo^{\eta}&  f      &\dTo_{\eta'} \\
\bullet&\rTo & \bullet  \\
\, &\xi'  &\, \\
\end{diagram}
}
\end{equation}
so that given for example $\xi$ and $\xi'$ the side path $\eta$ and $\eta'$ are uniquely determined{\footnote{Recall that for the fundamentals all the paths involved in the corresponding cells are of length $1$. }} thus determining $\rho$ and $\rho'$  to be equal to $\xi$ and $\xi'$. The constant of proportionality being the modulus of the cell.
The induction hypothesis is,
\begin{equation}
\label{dB6}
<\rho \otimes \rho'|\xi \otimes \xi'>_{f_1 \otimes\cdots \otimes f_{n-1}} =
Tr_{f_1 \otimes\cdots \otimes f_{n-1}} [(\Phi^{f_1 \otimes\cdots \otimes f_{n-1}} (\rho \otimes \rho'))^{\dagger} \Phi^{f_1 \otimes\cdots \otimes f_{n-1}} (\xi \otimes \xi')] =\delta_{\rho \xi}  \delta_{\rho' \xi'}
\end{equation}
we prove it for  
the case of $R= f_1 \otimes\cdots \otimes f_{n}$ we have,
\begin{eqnarray}
%\label{d6}
&&<\rho \otimes \rho'|\xi \otimes \xi'>_{f_1 \otimes\cdots \otimes f_{n}}
=
Tr_{f_1 \otimes\cdots \otimes f_{n}} [(\Phi^{f_1 \otimes\cdots \otimes f_{n}}(\rho \otimes \rho'))^{\dagger} 
\Phi^{f_1 \otimes\cdots \otimes f_{n}} (\xi \otimes \xi')] \nonumber\\
&=&\sum_{\lambda, \omega}
Tr_{f_1 \otimes\cdots \otimes f_{n}}[ 
\Phi^{f_1 \otimes\cdots \otimes f_{n-1}}(\lambda \otimes \rho')
(\Phi^{f_n} (\rho \otimes \lambda))^{\dagger}
\Phi^{f_1 \otimes\cdots \otimes f_{n-1}}(\omega \otimes \xi')
\Phi^{f_n} (\xi \otimes \omega)]\nonumber\\
&=& \sum_{\lambda, \omega}
Tr_{f_n} [(\Phi^{f_n} (\rho \otimes \lambda))^{\dagger} \Phi^{f_n} (\xi \otimes \omega)]
Tr_{f_1 \otimes\cdots \otimes f_{n-1}} [(\Phi^{f_1 \otimes\cdots \otimes f_{n-1}} (\lambda \otimes \rho'))^{\dagger} \Phi^{f_1 \otimes\cdots \otimes f_{n-1}} (\omega \otimes \xi')] \nonumber\\
&=& \sum_{\lambda, \omega} \delta_{\rho \xi} \delta_{\lambda \omega}
\delta_{\rho' \xi'}
\propto\delta_{\rho \xi}  \delta_{\rho' \xi'}
\end{eqnarray}
\end{proof}
Now since any representation appears in the decomposition of tensor products of the fundamentals, 
proposition \ref{d} implies proposition \ref{p2.3}.

%%%%%%%%%%%%%%%%%%%%%%%%%%%%%%%%%%%%%%%%%%%%%%%%%%%%%%%%%%%%%%%
%%%%%%%%%%%%%%%%%%%%%%%%%%%%%%%%%%%%%%%%%%%%%%%%%%%%%%%%%%%%%%%%%
%%%%%%%%%%%%%%%%%%%%%%%%%%%%%%%%%%%%%%%%%%%%%%%%%%%%%%%%%%%%%%

\section*{Appendix C. Morphism property of the coproduct.}

\begin{proposition}
\begin{equation}
\label{c1}
\Phi^{x_1}_{\alpha_1 \beta_1} \otimes \Phi^{x_2}_{\alpha_2 \beta_2}
[\Delta((\xi_1 \otimes \xi'_1)\cdot (\xi_2 \otimes \xi'_2))]
= 
\Phi^{x_1}_{\alpha_1 \beta_1} \otimes \Phi^{x_2}_{\alpha_2 \beta_2}
[\Delta((\xi_1 \otimes \xi'_1)) \cdot
\Delta((\xi_2 \otimes \xi'_2))]
\end{equation}
\end{proposition}
\begin{proof}
We split the proof in some steps,
\begin{enumerate}
\item [(i)]
\begin{equation}
\label{c2}
\Phi^{x_1}_{\alpha_1 \beta_1} \otimes \Phi^{x_2}_{\alpha_2 \beta_2}
\left( \Delta((\xi_1 \otimes \xi'_1)) \cdot
\Delta((\xi_2 \otimes \xi'_2)) \right)=
\Phi^{x_1}_{\alpha_1 \beta_1} \otimes \Phi^{x_2}_{\alpha_2 \beta_2}
\left( \Delta_{\mathcal P}((\xi_1 \otimes \xi'_1)) \star
\Delta_{\mathcal P}((\xi_2 \otimes \xi'_2)) \right)
\end{equation}
\begin{proof}
\begin{eqnarray}
\label{c3}
\Phi^{x_1}_{\alpha_1 \beta_1} \otimes \Phi^{x_2}_{\alpha_2 \beta_2}
\left( \Delta((\xi_1 \otimes \xi'_1)) \cdot
\Delta((\xi_2 \otimes \xi'_2)) \right)
&=&\sum_{\xi_i , \xi_j} 
\Phi^{x_1}_{\alpha_1 \beta_1} \otimes \Phi^{x_2}_{\alpha_2 \beta_2}
(\xi_1 \otimes \xi^i)_{\mathcal E}\otimes
(\xi_i \otimes \xi'_1)_{\mathcal E}\cdot
(\xi_2 \otimes \xi^j)_{\mathcal E}\otimes
(\xi_j \otimes \xi'_2)_{\mathcal E}
\nonumber\\
& =&\sum_{\xi_i , \xi_j}
\Phi^{x_1}_{\alpha_1 \beta_1}
((\xi_1 \otimes \xi^i)_{\mathcal E} \cdot
(\xi_2 \otimes \xi^j)_{\mathcal E})
\Phi^{x_2}_{\alpha_2 \beta_2}
((\xi_i \otimes \xi'_1)_{\mathcal E}\cdot
(\xi_j \otimes \xi'_2)_{\mathcal E})
\nonumber\\
& =&\sum_{\xi_i , \xi_j}
\Phi^{x_1}_{\alpha_1 \beta_1}
((\xi_1 \otimes \xi^i)_{\mathcal P} \star
(\xi_2 \otimes \xi^j)_{\mathcal P})
\Phi^{x_2}_{\alpha_2 \beta_2}
((\xi_i \otimes \xi'_1)_{\mathcal P}\star
(\xi_j \otimes \xi'_2)_{\mathcal P})
\nonumber\\
&=&
\Phi^{x_1}_{\alpha_1 \beta_1} \otimes \Phi^{x_2}_{\alpha_2 \beta_2}
\left( \Delta_{\mathcal P}((\xi_1 \otimes \xi'_1)) \star
\Delta_{\mathcal P}((\xi_2 \otimes \xi'_2)) \right)
\end{eqnarray}
where we have defined $\Delta_{\mathcal P}$ by,
\begin{equation}
\label{c4}
\Delta_{\mathcal P} (\xi \otimes \xi') = 
\sum_{\xi_i}
(\xi \otimes \xi^i)_{\mathcal P} \otimes(\xi_i \otimes \xi')_{\mathcal P}
\end{equation}
\end{proof}
\item [(ii)]
\begin{equation}
\label{c5}
 \Delta_{\mathcal P}((\xi_1 \otimes \xi'_1)) \star
\Delta_{\mathcal P}((\xi_2 \otimes \xi'_2)) 
=
 \Delta_{\mathcal P}((\xi_1 \otimes \xi'_1) \star
(\xi_2 \otimes \xi'_2)) 
\end{equation}
\begin{proof}
\begin{eqnarray}
\label{c6}
\Delta_{\mathcal P}((\xi_1 \otimes \xi'_1)) \star
\Delta_{\mathcal P}((\xi_2 \otimes \xi'_2)) 
&=&
\sum_{\xi_i , \xi_j, \#\xi_i = \#\xi_1 , \#\xi_j = \#\xi_2}
[(\xi_1 \otimes \xi^i)_{\mathcal P}\otimes
(\xi_i \otimes \xi'_1)_{\mathcal P}]\star
[(\xi_2 \otimes \xi^j)_{\mathcal P}\otimes
(\xi_j \otimes \xi'_2)_{\mathcal P}]
\nonumber\\
&=&
\sum_{\xi_i , \xi_j ,\#\xi_i = \#\xi_1 , \#\xi_j = \#\xi_2}
[(\xi_1 \star \xi_2)\otimes (\xi^i \star \xi^j)]_{\mathcal P}
\otimes
[(\xi_i \star \xi_j)\otimes (\xi'_1 \star \xi'_2)]_{\mathcal P}
\nonumber\\
&=&\Delta_{\mathcal P}((\xi_1 \otimes \xi'_1) \star
(\xi_2 \otimes \xi'_2))
\end{eqnarray}
where in the last equality we have used that a summation over 
elementary paths of length $n$ with fixed extreme vertex is equal to the summation over
any pair of paths of length $n_1$ and $n_2$ of the concatenation of them if $n_1 + n_2 = n$ and the starting(ending) vertex  of the first(last)path coincide with the ones of the length $n$ path.
\end{proof}
Finally we have,
\item [(iii)]
\begin{equation}
\label{c7}
\Phi^{x_1}_{\alpha_1 \beta_1} \otimes \Phi^{x_2}_{\alpha_2 \beta_2}
( \Delta_{\mathcal P}((\xi_1 \otimes \xi'_1) \star
(\xi_2 \otimes \xi'_2)) )
=
\Phi^{x_1}_{\alpha_1 \beta_1} \otimes \Phi^{x_2}_{\alpha_2 \beta_2}
(\Delta((\xi_1 \otimes \xi'_1) \cdot
(\xi_2 \otimes \xi'_2)) )
\end{equation}
\begin{proof}
\begin{eqnarray}
\label{c8}
\Phi^{x_1}_{\alpha_1 \beta_1} \otimes \Phi^{x_2}_{\alpha_2 \beta_2}
( \Delta_{\mathcal P}((\xi_1 \otimes \xi'_1) \star
(\xi_2 \otimes \xi'_2)) )&=&
\Phi^{x_1}_{\alpha_1 \beta_1} \otimes \Phi^{x_2}_{\alpha_2 \beta_2}
( \Delta_{\mathcal P}
\sum_{x,\alpha,\beta}                 
{
\begin{diagram}[size=0.8em,abut]
\,&\xi_1 \star \xi_2  &\, \\
\bullet&\rTo~{n} & \bullet  \\
\dTo^{\alpha}&  f      &\dTo_{\beta} \\
\bullet&\rTo & \bullet  \\
\, &\xi'_1 \star \xi'_2 &\, \\
\end{diagram}
}\;
E^x_{\alpha\beta})\nonumber\\
&=&
\Phi^{x_1}_{\alpha_1 \beta_1} \otimes \Phi^{x_2}_{\alpha_2 \beta_2}
( 
\sum_{x,\alpha,\beta, \xi, \xi'}                 
{
\begin{diagram}[size=0.8em,abut]
\,&\xi_1 \star \xi_2  &\, \\
\bullet&\rTo~{n} & \bullet  \\
\dTo^{\alpha}&  x      &\dTo_{\beta} \\
\bullet&\rTo & \bullet  \\
\, &\xi'_1 \star \xi'_2 &\, \\
\end{diagram}
}\;
{
\begin{diagram}[size=0.8em,abut]
\,&\xi  &\, \\
\bullet&\rTo~{n} & \bullet  \\
\dTo^{\alpha}&  x^{-1}      &\dTo_{\beta} \\
\bullet&\rTo & \bullet  \\
\, & \xi' &\, \\
\end{diagram}
}
\Delta_{\mathcal P}(\xi \otimes \xi' )
)
\nonumber\\
&=&
\sum_{x,\alpha,\beta,\xi, \xi',\xi_i}                 
{
\begin{diagram}[size=0.8em,abut]
\,&\xi_1 \star \xi_2  &\, \\
\bullet&\rTo~{n} & \bullet  \\
\dTo^{\alpha}&  x      &\dTo_{\beta} \\
\bullet&\rTo & \bullet  \\
\, &\xi'_1 \star \xi'_2 &\, \\
\end{diagram}
}\;
{
\begin{diagram}[size=0.8em,abut]
\,&\xi  &\, \\
\bullet&\rTo~{n} & \bullet  \\
\dTo^{\alpha}&  x^{-1}      &\dTo_{\beta} \\
\bullet&\rTo & \bullet  \\
\, &\xi' &\, \\
\end{diagram}
}
\Phi^{x_1}_{\alpha_1 \beta_1}(\xi \otimes \xi^i) 
\Phi^{x_2}_{\alpha_2 \beta_2}(\xi_i \otimes \xi')
\nonumber\\
&=&\sum_{\xi, \xi',\xi_i}                 
P^{\xi_1 \xi_2 \xi}_{\xi'_1 \xi'_2 \xi'}
{
\begin{diagram}[size=0.8em,abut]
\,&\xi  &\, \\
\bullet&\rTo~{n} & \bullet  \\
\dTo^{\alpha_1}&  x_{1}      &\dTo_{\beta_1} \\
\bullet&\rTo & \bullet  \\
\, &\xi^i &\, \\
\end{diagram}
}
{
\begin{diagram}[size=0.8em,abut]
\,&\xi_i  &\, \\
\bullet&\rTo~{n} & \bullet  \\
\dTo^{\alpha_2}&  x_{2}      &\dTo_{\beta_2} \\
\bullet&\rTo & \bullet  \\
\, &\xi' &\, \\
\end{diagram}
}
\nonumber\\
&=&
\Phi^{x_1}_{\alpha_1 \beta_1} \otimes \Phi^{x_2}_{\alpha_2 \beta_2}
(\Delta((\xi_1 \otimes \xi'_1) \cdot
(\xi_2 \otimes \xi'_2)) )
\end{eqnarray}
\end{proof}
\end{enumerate}
\end{proof}
%%%%%%%%%%%%%%%%%%%%%%%%%%%%%%%%%%%%%%%%%%%%%%%%%%%%%%%%%%%%%%%%%%%%%%%%%%%%%%%%%%%%
\section*{Appendix D. The trivial representation $\mathbf 0$.}
Consider a map
$\Phi^{\mathbf 0}_{\alpha_0 \beta_0} : End^{gr}(\mathcal E) \to {\mathbb C}$
defined such that for any irrep $x$ of the DTA and indices $\alpha_x , \beta_x$ in $x$
there exist unique subscripts
$\alpha_0 \beta_0$ in $\mathbf 0$ such that, 
\begin{equation}
\label{d11}
\Phi^{{\mathbf 0} \otimes x}_{\alpha_0 \star \alpha_x ,\beta_0 \star \beta_x}
(\xi \otimes \xi')=
\Phi^{x}_{\alpha_x , \beta_x}
(\xi \otimes \xi')
\end{equation}
using eqs. (\ref{2.21}) and (\ref{1.4}), eq. (\ref{d11}) implies,
\begin{equation}
\label{d22}
\sum_{\lambda}
{
\begin{diagram}[size=0.8em,abut]
\,&\xi  &\, \\
\bullet&\rTo & \bullet  \\
\dTo^{\alpha_0}&  {\mathbf 0}     &\dTo_{\beta_0} \\
\bullet&\rTo & \bullet  \\
\, &\lambda  &\, \\
\end{diagram}
}
\;
{
\begin{diagram}[size=0.8em,abut]
\,&\lambda  &\, \\
\bullet&\rTo & \bullet  \\
\dTo^{\alpha_x}&   x     &\dTo_{\beta_x} \\
\bullet&\rTo & \bullet  \\
\, &\xi'  &\, \\
\end{diagram}
}
=
{
\begin{diagram}[size=0.8em,abut]
\,&\xi  &\, \\
\bullet&\rTo & \bullet  \\
\dTo^{\alpha_x}&  x     &\dTo_{\beta_x} \\
\bullet&\rTo & \bullet  \\
\, &\xi'  &\, \\
\end{diagram}
}
\end{equation}
multiplying this last equation by,
$$
{
\begin{diagram}[size=0.8em,abut]
\,&\rho  &\, \\
\bullet&\rTo & \bullet  \\
\dTo^{\alpha_x}&   x^{-1}     &\dTo_{\beta_x} \\
\bullet&\rTo & \bullet  \\
\, &\rho'  &\, \\
\end{diagram}
}
$$
summing up over $x, \alpha_x, \beta_x$ and using (\ref{2.15'}) we obtain,
\begin{equation}
\label{d3}
{
\begin{diagram}[size=0.8em,abut]
\,&\xi  &\, \\
\bullet&\rTo & \bullet  \\
\dTo^{\alpha_0}&  {\mathbf 0}     &\dTo_{\beta_0} \\
\bullet&\rTo & \bullet  \\
\, &\xi'  &\, \\
\end{diagram}
}
= \delta_{\xi \xi'}
\end{equation}
Note that this last equation implies that
for $\xi_1,\xi'_1,\xi_2,\xi'_2$ such that $\#\xi_1 = \#\xi'_1 $,
$\#\xi_2 = \#\xi'_2 $ and $r(\xi_1 ) = s(\xi_2 )$, $r(\xi'_1 ) = s(\xi'_2 )$
we have,
\begin{equation}
\label{d4}
{
\begin{diagram}[size=0.8em,abut]
\,&\xi_1 \star \xi_2  &\, \\
\bullet&\rTo & \bullet  \\
\dTo^{\alpha_0}&  {\mathbf 0}     &\dTo_{\beta_0} \\
\bullet&\rTo & \bullet  \\
\, &\xi'_1 \star \xi'_2  &\, \\
\end{diagram}
}
=
{
\begin{diagram}[size=0.8em,abut]
\,&\xi_1   &\, \\
\bullet&\rTo & \bullet  \\
\dTo^{\alpha_0}&  {\mathbf 0}     &\dTo_{\gamma_0} \\
\bullet&\rTo & \bullet  \\
\, &\xi'_1   &\, \\
\end{diagram}
}
\;
{
\begin{diagram}[size=0.8em,abut]
\,&\xi_2   &\, \\
\bullet&\rTo & \bullet  \\
\dTo^{\gamma_0}&  {\mathbf 0}     &\dTo_{\beta_0} \\
\bullet&\rTo & \bullet  \\
\, &\xi'_2   &\, \\
\end{diagram}
}
\end{equation}
To show that this is indeed a representation  
note that eq.(\ref{2.15'''}) implies,
\begin{equation}
\label{d5}
\sum_{\xi , \xi'}
{
\begin{diagram}[size=0.8em,abut]
\,&\xi   &\, \\
\bullet&\rTo & \bullet  \\
\dTo^{\alpha}&  {\mathbf 0}^{-1}     &\dTo_{\beta} \\
\bullet&\rTo & \bullet  \\
\, &\xi'   &\, \\
\end{diagram}
}\;
{
\begin{diagram}[size=0.8em,abut]
\,&\xi   &\, \\
\bullet&\rTo & \bullet  \\
\dTo^{\alpha_0}&  {\mathbf 0}     &\dTo_{\beta_0} \\
\bullet&\rTo & \bullet  \\
\, &\xi'   &\, \\
\end{diagram}
}
=
\sum_{\xi}
{
\begin{diagram}[size=0.8em,abut]
\,&\xi   &\, \\
\bullet&\rTo & \bullet  \\
\dTo^{\alpha}&  {\mathbf 0}^{-1}     &\dTo_{\beta} \\
\bullet&\rTo & \bullet  \\
\, &\xi   &\, \\
\end{diagram}
}\;
=
\delta_{\alpha \alpha_0}\delta_{\beta \beta_0}
\end{equation}
hence,
\begin{equation}
\label{d6}
\Phi^{\mathbf 0}_{\alpha_0 \beta_0}(E^{\mathbf 0}_{\alpha \beta})=
\sum_{\xi \xi'}
{
\begin{diagram}[size=0.8em,abut]
\,&\xi   &\, \\
\bullet&\rTo & \bullet  \\
\dTo^{\alpha}&  {\mathbf 0}^{-1}     &\dTo_{\beta} \\
\bullet&\rTo & \bullet  \\
\, &\xi'   &\, \\
\end{diagram}
}\;
\Phi^{\mathbf 0}_{\alpha_0 \beta_0}(\xi \otimes \xi')=
\sum_{\xi}
{
\begin{diagram}[size=0.8em,abut]
\,&\xi   &\, \\
\bullet&\rTo & \bullet  \\
\dTo^{\alpha}&  {\mathbf 0}^{-1}     &\dTo_{\beta} \\
\bullet&\rTo & \bullet  \\
\, &\xi   &\, \\
\end{diagram}
}\;
=
\delta_{\alpha \alpha_0} \delta_{\beta \beta_0}
\end{equation}
hence $\Phi^{\mathbf 0}_{\alpha_0 \beta_0}(E^{\mathbf 0}_{\alpha_0 \beta_0})$
is of the form (\ref{2.6}) thus being a homomorphism for the matrix product
to which the product in the DTA reduces in this basis.

%%%%%%%%%%%%%%%%%%%%%%%%%%%%%%%%%%%%%%%%%%%%%%%%%%%%%%%%%%%%%%%%%

\end{document}